\documentclass[twoside,11pt]{article}

\usepackage{blindtext}

\usepackage[nohyperref, preprint]{jmlr2e}

\usepackage{tabularx} 
\usepackage{amsmath}  
\usepackage{mathrsfs}
\usepackage{enumitem}
\usepackage{xcolor}
\usepackage{amsfonts}
\usepackage{algorithm}
\usepackage{algorithmic}
\usepackage{soul}
\usepackage{titletoc}

\usepackage{hyperref}
\definecolor{darkpowderblue}{rgb}{0.0, 0.2, 0.6}
\definecolor{goldenpoppy}{rgb}{0.99, 0.76, 0.0}
\definecolor{cardinal}{rgb}{0.77, 0.12, 0.23}
\hypersetup{
    colorlinks=true,
    linkcolor=cardinal,
    citecolor=darkpowderblue,      
    urlcolor=cyan,
}

\newtheorem{assumption}{Assumption}
\newtheorem{condition}{Condition}

\newcommand{\F}{G}
\newcommand{\f}{g}
\newcommand{\Ld}{H}
\newcommand{\ld}{h}
\renewcommand{\L}{L}
\newcommand{\R}{R}
\newcommand{\Tt}{\Theta}

\newcommand{\Dt}{\Delta}
\newcommand{\ap}{\alpha}

\newcommand{\tLd}{\tilde{\Ld}}

\newcommand{\tF}{\tilde{\F}}

\newcommand{\tL}{\tilde{\L}}
\newcommand{\tR}{\tilde{\R}}

\newcommand{\tap}{\tilde{\ap}}

\newcommand{\hLd}{\hat{\Ld}}

\newcommand{\hF}{\hat{\F}}

\newcommand{\hL}{\hat{\L}}
\newcommand{\hR}{\hat{\R}}

\newcommand{\hap}{\hat{\ap}}
\newcommand{\hTt}{\hat{\Tt}}

\DeclareMathOperator{\myspan}{span}

\DeclareMathOperator{\rank}{rank}

\DeclareMathOperator{\fn}{F}
\DeclareMathOperator{\opn}{op}

\DeclareMathOperator*{\argmin}{arg\,min}

\allowdisplaybreaks

\usepackage{lastpage}
\jmlrheading{xx}{xxxx}{1-\pageref{LastPage}}{x/xx; Revised x/xx}{x/xx}{xx-xxxx}{Xuanyu Chen, Xiaoyue Niu, Gongjun Xu and Ji Zhu}

\makeatletter
\let\originalpsjmlrtps\ps@jmlrtps
\def\ps@jmlrtps{%
  \originalpsjmlrtps
  \def\@oddfoot{}%
  \def\@evenfoot{}%
}
\makeatother

\ShortHeadings{Noisy Matrix Completion under Informative Missingness}{Chen, Niu, Xu and Zhu}
\firstpageno{1}

\begin{document}

\title{Noisy Matrix Completion under Informative Missingness}

\author{\name Xuanyu Chen \email xuanyuch@umich.edu \\
       \addr Department of Statistics\\
       University of Michigan\\
       Ann Arbor, MI 48109-1107, USA
       \AND
       \name Xiaoyue Niu \email xiaoyue@psu.edu \\
       \addr Department of Statistics\\
       Pennsylvania State University\\
       University Park, PA 16802, USA
       \AND
       \name Gongjun Xu \email gongjun@umich.edu \\
       \addr Department of Statistics\\
       University of Michigan\\
       Ann Arbor, MI 48109-1107, USA
       \AND
       \name Ji Zhu \email jizhu@umich.edu \\
       \addr Department of Statistics\\
       University of Michigan\\
       Ann Arbor, MI 48109-1107, USA
       }

\editor{My editor}

\maketitle

\begin{abstract}

\noindent
Noisy matrix completion is a fundamental problem in statistical learning and has attracted a substantial amount of interest over the last two decades. In a variety of applications, the missingness pattern is highly informative, yet it has received relatively less attention. Most existing methods either overlook this source of information or oversimplify its generating process, with few attempting to model the association between the data matrix and its informative missingness. In this study, we propose a general modeling framework that jointly models the data matrix and its missingness pattern using a pair of low-rank models coupled by a flexible shared linear structure. This joint low-rank approach incorporates the informative missingness to improve matrix completion and can flexibly capture various associations between the two modes of data. We develop an efficient joint estimation procedure for this framework based on projected gradient descent, and establish local convergence guarantees that unveil its computational and statistical properties. We further demonstrate, through extensive simulation studies and real-world data analysis, that our proposed approach outperforms competing methods, achieving substantial improvements in prediction accuracy on missing entries.

\end{abstract}


\section{Introduction}\label{sec:intro}

Noisy matrix completion is the task of completing and denoising a typically large-scale \textit{target matrix} from an observed \textit{data matrix} with incomplete and noisy entries. 
This problem naturally arises in a wide range of applications, including recommender systems 
\citep{rennie2005fast, koren2009matrix, bobadilla2013recommender, dai2019recommender}, 
image processing \citep{ji2010robust, luo2015multiview}, survey imputation 
\citep{zhang2020active, sengupta2021simple, mao2024mixed}, 
causal inference \citep{athey2021matrix, choi2024matrix}, among others. It has attracted significant interest from the statistical learning community over the last two decades. Most statistical methods are based on the assumption that the target matrix is of low rank 
\citep{candes2010matrix, cai2010singular, recht2011simpler, rohde2011estimation, negahban2012restricted, klopp2014, klopp2015, cai2016matrix, fithian2018flexible, mao2021matrix, ma2024statistical}, 
in order to address the high dimensionality of the data matrix, a fundamental challenge in matrix completion. Apart from the data matrix itself, in many applications of matrix completion, its missingness pattern is also a highly informative component of the data. 
For example, in movie rating data, 
the missingness pattern is usually indicative of user preferences, as users often tend to rate movies that align with their interests while leaving others unrated \citep{marlin2007}.
In this case, properly accounting for the informative missingness
may lead to improved reconstruction quality of the target matrix.

However, despite the popularity of matrix completion, the treatment of the missingness pattern has received considerably less attention. 
Most existing works either overlook this informative part of the data or oversimplify its generating process, which can be complicated in reality. 
Early studies \citep{candes2010matrix, recht2011simpler, rohde2011estimation} primarily consider uniform missingness, 
i.e., observed entries are uniformly sampled from the complete data matrix. While this idealized setting is often unrealistic in practice, its analysis 
lays the cornerstone for the theoretical advancements on this topic. 
Some subsequent works \citep{klopp2014, klopp2015, cai2016matrix} study noisy matrix completion under a general sampling distribution, where entries can be observed independently with different probabilities, without any adjustment for the non-uniform missingness in their loss function. 
Their analyses provide insight into the extent to which techniques developed for uniform missingness remain valid for more general cases, but this approach fails to make full use of the data. 
Several studies aim to improve the estimation of the target matrix via modeling the non-uniform missingness and adapting for it. 
For instance, \citet{srebro2010collaborative}, \citet{foygel2011learning}, and \citet{negahban2012restricted} adopt a product sampling model, which samples an entry by selecting a row and a column independently from potentially non-uniform marginal distributions. This model results in an observation probability matrix of rank one, and they adjust for it in their method by using certain weighted nuclear-norm regularizations. Nonetheless, a rank-one model on missingness is often too restrictive in many applications.

Methods attempting to incorporate the informative missingness with a flexible missingness model are relatively scarce. Inverse probability weighting (IPW) \citep{imbens2015causal} is a widely adopted approach.  
Under this framework, one typically assumes a model for the missingness pattern, estimates the observation probability matrix, and then uses these probabilities to debias the loss function.
Specifically, \citet{schnabel2016recommendations}, \citet{wang2018modeling}, \citet{wang2018collaborative}, \citet{wang2019doubly},
and \citet{ma2024statistical} assume regression models for the missingness pattern given predictors. \citet{ma2019missing} and \citet{mao2021matrix} suppose that it follows another non-linear low-rank model. \citet{bhattacharya2022matrix} model the observation probability matrix as an entry-wise transformation of the target matrix. 
However, in matrix completion, the observation rate is typically very low, leading to high variation and sensitivity to model misspecification of the IPW estimator.
Apart from IPW-based methods, there is a separate body of work focusing on modeling the statistical dependency between the data matrix and its missingness pattern 
\citep{hernandez2014probabilistic, sportisse2020imputation, jin2022matrix, li2024pairwise}.

Moreover, one common limitation of the above methods is that most of them fail to properly model the underlying association between the data matrix and its informative missingness. 
Motivated by these observations, in this paper, we propose a general framework that jointly models the data matrix and its missingness pattern using a pair of low-rank models, whose parameters are linked by certain shared linear structure. 
Our joint low-rank approach can be understood through the lens of latent space modeling. It assumes that the latent embeddings governing the data matrix and its missingness are both derived from a set of unified embeddings, and thus carry subsets of information that can flexibly overlap.
Compared to existing work, our approach has the following advantages. First, modeling the missingness pattern with a low-rank model is more general and can provide a more flexible fit to the observed pattern than assuming either uniform missingness \citep{candes2010matrix, recht2011simpler, rohde2011estimation} or a rank-one missingness model
\citep{srebro2010collaborative, foygel2011learning, negahban2012restricted}. 
Second, rather than directly applying methods designed for uniform missingness to the non-uniform case \citep{klopp2014, klopp2015, cai2016matrix}, our approach leverages shared parameters between the two sub-models to incorporate the informative missingness in the estimation of the target matrix. This also avoids the previously mentioned issue of using IPW. Third, a unique advantage of our proposed framework is that it can characterize various possible associations between the two modes of data explicitly and flexibly, by employing different shared linear structures in the parameters. In particular, it includes as special cases the two low-rank approaches briefly mentioned by \citet{fithian2018flexible} for jointly modeling the data matrix and its missingness. 
In the appendix, 
we briefly review related models employing shared latent structures for matrix completion and other problems.

The rest of the paper is organized as follows. We formally introduce the joint low-rank modeling framework in Section~\ref{sec:method}. In Section~\ref{sec:estimation}, we develop an efficient joint estimation procedure, initialized by separately estimating the two individual low-rank models. 
As the procedure involves non-convex optimization, in Section~\ref{sec:theory} we establish local convergence results that unveil its computational and statistical properties. 
In Section~\ref{sec:model_selection}, we discuss the selection of latent dimensions. 
Simulation studies are conducted in Section~\ref{sec:simulation}. 
Our real-world data analysis in Section~\ref{sec:real_data} demonstrates that the joint modeling approach achieves significantly higher prediction accuracy than competing methods on unseen entries and properly integrates the information from the two data sources. All proofs can be found in the appendix.

\section{A Joint Low-Rank Modeling Framework}\label{sec:method}

Let $Y \in \mathbb{R}^{n \times p}$ be the underlying complete data matrix, which is only partially observed. Its missingness pattern is encoded in a binary matrix $Z\in\{0,1\}^{n\times p}$, where $Z_{ij}=1$ if $Y_{ij}$ is observed, and $0$ otherwise. Abusing notation slightly, we denote the observed data by $\{Y, Z\}$.
We refer to $\mathbb{E}[Y]$ as the target matrix, whose estimation is the main goal in matrix completion.

\paragraph{Notation} For any matrix $A$, let $\|A\|_{\fn}$, $\|A\|_{\opn}$, $\|A\|_{*}$, $\|A\|_{2,\infty}$ and $\|A\|_{\infty}:= \max_{i,j}|A_{ij}|$ be its Frobenius, operator, nuclear, $2$-to-infinity and entry-wise infinity norms, respectively. $\sigma_+(A)$, $\sigma_-(A)$ and $\sigma_k(A)$ denote the largest, smallest and $k$-th largest singular values of $A$, respectively. Let $\myspan(A)$ be the column space of $A$. 
For any two matrices $A\in\mathbb{R}^{n\times p}$ and $B\in\mathbb{R}^{n\times q}$, $(A\ B)\in \mathbb{R}^{n\times (p+q)}$ denotes their concatenation by columns. If $A$ and $B$ have the same size, denote by $A \circ B$ their component-wise product. For a set of matrices $\{A_m\}_{m=1}^M$ indexed by $m$, let $A_{m,ij}$ be the $(i,j)$-entry of $A_m$. Let $\mathbb{O}^d$ be the set of all $d$-by-$d$ orthogonal matrices. For any $\mathcal{C}\subset \mathbb{R}^{n\times p}$, let $\mathcal{P}_\mathcal{C}(\cdot)$ be the projection operator onto $\mathcal{C}$ with respect to the Frobenius norm. 
For any vector $\alpha$, let $\|\alpha\|_{2}$ and $\|\alpha\|_{\infty}$ be its vector $2$-norm and infinity norm. For a set of vectors $\{\alpha_m\}_{m=1}^M$, denote the $i$-th entry of $\alpha_{m}$ by $\alpha_{m,i}$. For any $\mathcal{C}\subset \mathbb{R}^{d}$, let $\mathcal{P}_\mathcal{C}(\cdot)$ be the projection operator onto $\mathcal{C}$ with respect to vector $2$-norm. 
For $a,b\in\mathbb{R}$, $\lfloor a\rfloor$ denotes the greatest integer less than or equal to $a$, $a\vee b:= \max\{a,b\}$, and $a\wedge b:= \min\{a, b\}$. For two deterministic sequences $\{a_n\}$ and $\{b_n\}$, $b_n = \omega(a_n)$ if $a_n = o(b_n)$, and $b_n = \Omega(a_n)$ if $a_n = O(b_n)$. We use $a_n = O(b_n)$ and $a_n\lesssim b_n$ interchangeably. Let $o_P, O_P, \omega_P$, and $\Omega_P$ be their probabilistic counterparts. For $n\in\mathbb{Z}_+$ denote $[n]:= \{1,2,\cdots, n\}$.

\subsection{Individual Low-Rank Models}\label{sec:individual}

We assume that $Y$ follows a linear low-rank model, 
$$Y_{ij} = \f_{1,i}^\top \ld_{1,j} + \ap_{1,i} + E_{ij}\quad \text{for } i\in[n], j\in[p],$$
where $\f_{1,i}, \ld_{1,j}\in\mathbb{R}^{K_1}$ and $\ap_{1,i}\in\mathbb{R}$ are fixed parameters, $K_1\ll n\wedge p$ is the latent dimension, and $E_{ij}$'s are independently distributed random errors with mean zero. Under this formulation, $\f_{1,i}$ and $\ld_{1,j}$ can be viewed as latent embeddings of row $i$ and column $j$ in a $K_1$-dimensional latent space, explaining the variation in $Y_{ij}$ through their inner product. 
Specifically, the coordinates of $\f_{1,i}$ can be interpreted as the latent attributes of row $i$, and those of $\ld_{1,j}$ are the scoring of column $j$ on these attributes. $\ap_{1,i}$'s are the row-specific effects that capture their heterogeneity. 

To model the non-uniform missingness, with the hypothesis that the generation of missingness pattern $Z$ is also governed by a small set of latent attributes as is the data matrix $Y$, we assume that $Z$ follows another (non-linear) low-rank model, 
$$
Z_{ij}\overset{\mathrm{ind.}}{\sim} \mathrm{Bernoulli}(\sigma(\f_{2,i}^\top \ld_{2,j} + \ap_{2,i}))\quad\text{for } i\in[n], j\in[p].
$$
Here $\f_{2,i}, \ld_{2,j}\in\mathbb{R}^{K_2}\ (K_2\ll n\wedge p)$ and $\ap_{2,i}\in\mathbb{R}$ are again fixed parameters, and $\sigma(x) = (1+e^{-x})^{-1}$ is the sigmoid function. The interpretation of $\f_{2,i}, \ld_{2,j}$, and $\alpha_{2,i}$ mirrors that of $\f_{1,i}, \ld_{1,j}$, and $\alpha_{1,i}$, but is associated with the missingness pattern.

Stacking $\f_{m,i}$ and $\ld_{m,j}$ as rows into matrices $\F_m,\Ld_m\in\mathbb{R}^{n\times K_m}$ and letting $\ap_m = (\ap_{m,i})_{i=1}^n\in\mathbb{R}^n$ for $m=1,2$, we adopt the following matrix notations,
\begin{align}
Y&= \F_1\Ld_1^\top + \ap_1 1_{p}^\top + E,\label{eq:sep_model1}\\
Z&\sim \mathrm{Bernoulli}(\sigma(\F_2\Ld_2^\top + \ap_2 1_{p}^\top)),\label{eq:sep_model2}
\end{align}
in which $E$ is a noise matrix with independent, mean-zero entries, and $\sigma(\cdot)$ applies element-wise. Under this formulation, the parameter matrices $\mathbb{E}[Y]$ and $\mathrm{logit}(\mathbb{E}[Z])$ have ranks at most $K_1+1$ and $K_2+1$.
Here we intentionally leave the joint distribution of $E$ and $Z$ unspecified to maintain the flexibility of our modeling framework. We will introduce specific assumptions later when establishing the theoretical guarantee of our proposed estimation procedure in Section~\ref{sec:theory}. 
Note that it is possible to add column-specific effects (e.g., $\beta_1 \in \mathbb{R}^p$) to capture column heterogeneity, which would extend the model to
$\mathbb{E}[Y] = \F_1\Ld_1^\top + \ap_1 1_{p}^\top + 1_n \beta_1^\top$ (and similarly for $Z$). 
We omit such terms in our model to maintain its simplicity, while our estimation method and theoretical results can be naturally extended to accommodate them.

\subsection{The Shared Linear Structure}\label{sec:shared}

The missingness can be informative if the probability of observations $\mathbb{E}[Z]$ is related to the target matrix $\mathbb{E}[Y]$. 
In order to model such association between the data matrix $Y$ and its missingness pattern $Z$ with flexibility, we further assume that the underlying parameters of the two low-rank models are linked by the shared linear structure described below. 

We associate each row $i$ with a single \textit{unified embedding} $\f_i\in\mathbb{R}^r$, and each column $j$ with a unified embedding $\ld_j\in\mathbb{R}^s$, where the dimensions satisfy $r,s \geq K_1 \vee K_2$. Then we assume that embeddings $\{\f_{1,i}, \f_{2,i}\}$ of rows and $\{\ld_{1,j}, \ld_{2,j}\}$ of columns are linear transformations of their corresponding unified embeddings, $\f_i$ and $\ld_j$, respectively. Formally, we have
$$
\f_{m,i} = \L_m^\top \f_i \quad \text{and}\quad \ld_{m,j} = \R_m^\top \ld_j,
$$
for some full-rank coefficient matrices $\L_m\in\mathbb{R}^{r\times K_m}$, $\R_m\in\mathbb{R}^{s\times K_m}$ and $m=1,2$.  
The unified embeddings $\f_i$'s and $\ld_j$'s thus link the generation of the data matrix and its missingness pattern. 
Stacking $\f_i$'s and $\ld_j$'s as rows into matrices $\F \in\mathbb{R}^{n\times r}$ and $\Ld \in\mathbb{R}^{p\times s}$, 
our proposed joint low-rank modeling framework is parametrized as follows,
\begin{equation}\label{eq:model}
\begin{aligned}
Y&= \F_1\Ld_1^\top + \ap_1 1_{p}^\top + E,\\
Z&\sim \mathrm{Bernoulli}(\sigma(\F_2\Ld_2^\top + \ap_2 1_{p}^\top)),\quad \text{where }\\
\underset{n\times K_1}{\F_1} &= \underset{n\times r}{\F}{\L_1}, \underset{p\times K_1}{\Ld_1} = \underset{p\times s}{\Ld}{\R_1}, \text{ with } K_1\leq r,s,\\ 
\underset{n\times K_2}{\F_2} &= \underset{n\times r}{\F}{\L_2}, \underset{p\times K_2}{\Ld_2} = \underset{p\times s}{\Ld}{\R_2}, \text{ with } K_2\leq r,s. 
\end{aligned}
\end{equation}
We further require $r,s\leq K_1 + K_2$ and certain full-rank condition (Condition~\ref{ass:ident_rank}), so that $(\L_1\ \L_2)$ and $(\R_1\ \R_2)$ have full row rank and the parameters $\F$ and $\Ld$ are identifiable. Detailed discussion on the identifiability issue is deferred to Section~\ref{sec:ident}.
The introduced shared linear structure can be equivalently characterized by the following condition on the column spaces of $\F_1$, $\F_2$, $\Ld_1$ and $\Ld_2$, 
\begin{equation}\label{eq:common_subspace}
\begin{aligned}
 \dim\{\myspan(\F_1)\cap\myspan(\F_2)\} &= r_0\geq 0,\\
 \dim\{\myspan(\Ld_1)\cap\myspan(\Ld_2)\} &= s_0\geq 0,
\end{aligned}
\end{equation}
in which $r_0 = K_1 + K_2-r$ and $s_0 = K_1 + K_2-s$.

A key advantage of the proposed shared linear structure is that, with appropriately chosen $K_1$, $K_2$, $r$ and $s$, it can flexibly represent various associations between the data matrix and its missingness. 
Taking the row embeddings as an example, intuitively, $\f_{1,i}$ and $\f_{2,i}$ each contain a subset of the information encoded in the unified embedding $\f_{i}$. Their relationship is thus determined by how their respective information overlaps. 
When $r=K_1\vee K_2$, $\F_2$ encodes as much information as $\F$, since one can recover $\F$ given the knowledge of $\F_2$ and  $\L_2$. Assuming $K_1\leq K_2$ without loss of generality, as $\myspan(\F_1)\subseteq \myspan(\F_2)$, the information $\F_1$ carries is a subset of that carried by $\F_2$. When $K_1\vee K_2<r< K_1 + K_2$, the information contained in $\F_1$ and $\F_2$ has non-empty overlap as $\dim\{\myspan(\F_1)\cap \myspan(\F_2)\} = r_0>0$. Moreover, they both carry their unique information, as knowing either alone is insufficient to recover $\F$. 
We note that this scenario is often the case in practice.
In the movie rating example, apart from factors affecting both whether and how a user would rate, factors like a movie's cinematography quality influence only the rating itself, and there are also factors affecting only a user's decision to rate, such as a movie's availability on their streaming service.
Lastly, $r = K_1 + K_2$ is the most general case where we allow $\F_1$ and $\F_2$ to be free parameters, indicating no shared information. The same argument applies to column embeddings as well.

\subsection{Identifiability of the Joint Model}\label{sec:ident}

Our proposed joint model \eqref{eq:model} is not identifiable without additional conditions. The parameters suffer from two types of ambiguity: first, an additive ambiguity involving the intercept terms, and second, a multiplicative ambiguity due to matrix factorizations. These issues pose technical challenges for model estimation and its analysis. To resolve these ambiguities, we introduce the following identifiability condition on model~\eqref{eq:model}. 

\begin{condition}\label{ass:ident}
\begin{enumerate}[ref={\thecondition.\arabic*}, leftmargin=*]
\item \label{ass:ident_size} $n> r$, $p> s$, and $K_1 \vee K_2\leq r,s\leq K_1+K_2$;
\item \label{ass:ident_rank} All parameter matrices are of full rank, i.e., $\rank(\F) = r$, $\rank(\Ld) = s$, $\rank(\L_1) = \rank(\R_1) = K_1$, $\rank(\L_2) = \rank(\R_2) = K_2$, $\rank((\L_1\ \L_2)) = r$ and $\rank((\R_1\ \R_2))=s$;
\item \label{ass:ident_cen} $\Ld$ is column-wise centered, i.e., $\Ld^\top 1_p = 0_s$;
\item \label{ass:ident_prod} It holds that
\begin{align*}
\F^\top \F = nI_r,&\ \Ld^\top\Ld =pI_s,\\
\L_1^\top \L_1=\R_1^\top \R_1,&\ \L_2^\top \L_2=\R_2^\top \R_2.
\end{align*}
\end{enumerate}
\end{condition}

Condition~\ref{ass:ident_rank} essentially requires that there be no redundant columns in $\F$, $\F_1$, $\F_2$, $\Ld$, $\Ld_1$ or $\Ld_2$, i.e., the role of each column in generating the data cannot be replaced by the other columns in the same matrix. Otherwise, the model can be equivalently parametrized with lower latent dimensions so that Condition~\ref{ass:ident_rank} is satisfied.
Conditions~\ref{ass:ident_cen} and \ref{ass:ident_prod} present one possible solution to resolve 
the additive and multiplicative ambiguities, respectively. 
Intuitively, the first two equations in Condition~\ref{ass:ident_prod} require that the unified latent attributes (columns of $\F$ and $\Ld$) be orthogonal and of the same magnitude, which could potentially minimize the information overlap between them. The last two constraints enforce a structural equivalence between the row and the column transformations, ensuring that their contributions to both sub-models are balanced.
Under Condition~\ref{ass:ident}, the parameters in model~\eqref{eq:model} are identifiable up to certain rotations, as detailed in the following proposition. 
 
\begin{proposition}\label{prop:ident}

Suppose that Condition~\ref{ass:ident} holds for two sets of parameters of model~\eqref{eq:model}, $\Xi = \{\F, \Ld, \L_1, \L_2, \R_1, \R_2, \ap_1, \ap_2\}$ and $\tilde{\Xi}$. Then they specify the same distribution of data, 
if and only if $\ap_1=\tilde{\ap}_1$, $\ap_2=\tilde{\ap}_2$ and there exist rotations $O_\F\in\mathbb{O}^r$, $O_1\in\mathbb{O}^{K_1}$, $O_2\in\mathbb{O}^{K_2}$ and $O_\Ld\in\mathbb{O}^s$ such that
\begin{align*}
\F\L_1\R_1^\top\Ld^\top &= \underbrace{(\F{O_\F})}_{=\tilde{\F}}
\underbrace{(O_\F^\top \L_1 {O_1})}_{=\tilde{\L}_1}
\underbrace{(O_1^\top \R_1^\top O_\Ld)}_{=\tilde{\R}_1^\top} 
\underbrace{(O_\Ld^{\top}\Ld^\top)}_{=\tilde{\Ld}^\top}, \\
\F\L_2\R_2^\top\Ld^\top &= (\F{O_\F})
\underbrace{(O_\F^\top \L_2 {O_2})}_{=\tilde{\L}_2}
\underbrace{(O_2^\top \R_2^\top O_\Ld)}_{=\tilde{\R}_2^\top} 
(O_\Ld^{\top}\Ld^\top).
\end{align*}
\end{proposition}

We note that in some tasks where the interpretation of each latent attribute is the primary concern, the remaining rotational indeterminacy needs to be further addressed. With a fitted model, this can be done, for example, by finding rotations that satisfy certain constraints or minimize a certain complexity measure, so that the rotated latent embeddings exhibit a simpler pattern (e.g., sparsity) and thus have better interpretability \citep{browne2001overview, liu2023rotation, zeng2023vintage}.

\section{Model Estimation}\label{sec:estimation}

In this section, we present the estimation procedure of the joint low-rank model~\eqref{eq:model}. We propose a joint estimation approach, in which the two sub-models are jointly estimated by projected gradient descent over the factorized parameter matrices, a natural choice in view of the shared linear structure. This approach is computationally efficient, but coming at the cost of non-convexity. We initialize it by separately fitting the low-rank models on $Y$ and $Z$. Note that to apply the proposed procedure, the latent dimensions $K_1$, $K_2$, $r$ and $s$ need to be properly determined. We discuss the selection of them in Section~\ref{sec:model_selection}.

Let $\Tt_m =\F_m\Ld_m^\top + \ap_m 1_p^\top = \F\L_m\R_m^\top \Ld^\top + \ap_m 1_p^\top$ for $m=1,2$. For the partially observed data matrix $Y$, we consider the following least-squares loss function
\begin{align}\label{eq:sep_loss1}
L_{1}(\Tt_1) = \frac{1}{2}\sum_{i,j}Z_{ij}(Y_{ij} - \Tt_{1,ij})^2 = \frac{1}{2}\|Z\circ (Y-\Tt_1)\|_{\fn}^2,  
\end{align}
and for the missingness pattern $Z$, we use its negative log-likelihood function
\begin{align}\label{eq:sep_loss2}
L_{2}(\Tt_2) = -\sum_{i,j}\left\{Z_{i j} \Tt_{2,i j}+\log \left(1-\sigma\left(\Tt_{2,i j}\right)\right)\right\}. 
\end{align}

\subsection{Joint Estimation}\label{sec:joint_est}

Denote by $\Xi = \{\F, \Ld, \L_1, \L_2, \R_1, \R_2, \ap_1, \ap_2\}$ the set of all parameters. To estimate the joint low-rank model~\eqref{eq:model}, we consider the following penalized combined loss function 
\begin{align}\label{eq:combined_loss}
\mathcal{L}(\Xi) := w {L}_{1}(\Tt_1) + (1-w) L_{2}(\Tt_2)+ np P(\Xi),
\end{align}
with a weight tuning parameter $w\in (0,1)$ and a penalty term
$$
P(\Xi) = \gamma_1 \Big\|\frac{\F^\top \F}{n}-I_r\Big\|_{\fn}^2 
+ \gamma_1 \Big\|\frac{\Ld^\top \Ld}{p}-I_s\Big\|_{\fn}^2
+ \gamma_2 \|\L_1^\top \L_1-\R_1^\top \R_1\|_{\fn}^2 
+ \gamma_2 \|\L_2^\top \L_2-\R_2^\top \R_2\|_{\fn}^2.
$$
The weight $w$ reflects the relative importance of the two sub-models in the joint model. Without prior information, one may simply set $w=0.5$. 
The penalty term $P(\Xi)$, instead of regularizing the estimation, enforces Condition~\ref{ass:ident_prod} without harming the modeling capacity. 
To see this, keeping the value of $w {L}_{1}(\Tt_1) + (1-w) L_{2}(\Tt_2)$
unchanged, we can always find a set of parameters satisfying Condition~\ref{ass:ident_prod} so that the penalty term is $0$. 
Note that although $P(\Xi)$ takes a rather complicated form, it is differentiable and the evaluation of it and its derivative has computational complexity of lower order. We explain the rationale behind the penalty later in Lemma~\ref{lemma:equiv_metric_main} and the subsequent discussion. 
For the two tuning parameters $\gamma_1$ and $\gamma_2$ in $P(\Xi)$, our theoretical analysis in Section~\ref{sec:eb_joint_est} provides a practical guideline for determining their ratio, which reduces the problem to tuning a single parameter, the overall penalty strength.

We minimize the combined loss $\mathcal{L}(\Xi)$ using projected gradient descent, as summarized in Algorithm~\ref{alg:joint_est}. 
In each iteration, the algorithm updates each parameter in $\Xi$ by first moving along the negative gradient direction by a pre-specified step size, and then projecting the intermediate estimate onto a pre-defined constraint set. 

\begin{algorithm}[htb!]
\caption{Projected gradient descent for the joint estimation}\label{alg:joint_est}
\begin{algorithmic}[1]
\STATE \textbf{Input:} data $\{Y,Z\}$; initialization $\Xi^0$; penalty parameters $\gamma_1$, $\gamma_2$; step sizes $\eta_\F$, $\eta_\Ld$, $\eta_{\L_1}$, $\eta_{\L_2}$, $\eta_{\R_1}$, $\eta_{\R_2}$, $\eta_{\ap_1}$, $\eta_{\ap_2}$; constants $M_0$, $M_1$, $M_2$ defining the projection sets; number of iterations $T$;

\FOR{$t=0,\ldots, T-1$}

\STATE $\tF^{t+1} = \F^{t} - \eta_\F \nabla_{\F}\mathcal{L}(\Xi^t)$;
\STATE $\tLd^{t+1} = \Ld^{t} - \eta_\Ld \nabla_{\Ld}\mathcal{L}(\Xi^t)$;

\STATE $\tL_m^{t+1} = \L_m^{t} - \eta_{\L_m} \nabla_{\L_m}\mathcal{L}(\Xi^t)$ for $m=1,2$;
\STATE $\tR_m^{t+1} = \R_m^{t} - \eta_{\R_m} \nabla_{\R_m}\mathcal{L}(\Xi^t)$ for $m=1,2$;
\STATE $\tap_m^{t+1} = \ap_m^{t} - \eta_{\ap_m} \nabla_{\ap_m}\mathcal{L}(\Xi^t)$ for $m=1,2$;
\STATE $\F^{t+1} = \mathcal{P}_{\mathcal{C}(\F)}(\tF^{t+1})$, $\Ld^{t+1} = \mathcal{P}_{\mathcal{C}(\Ld)}(\tLd^{t+1})$, 
$\L_m^{t+1} = \mathcal{P}_{\mathcal{C}(\L_m)}(\tL_m^{t+1})$, \\
$\R_m^{t+1} = \mathcal{P}_{\mathcal{C}(\R_m)}(\tR_m^{t+1})$, $\ap_m^{t+1} = \mathcal{P}_{\mathcal{C}(\ap_m)}(\tap_m^{t+1})$ for $m=1,2$;
\STATE $\Xi^{t+1} = \{\F^{t+1}, \Ld^{t+1}, \L_1^{t+1}, \L_2^{t+1}, \R_1^{t+1}, \R_2^{t+1}, \ap_1^{t+1}, \ap_2^{t+1}\}$;
\ENDFOR
\STATE \textbf{Output:} $\hat{\Xi} = \Xi^T$.
\end{algorithmic}
\end{algorithm}

We note that our theoretical analysis in Section~\ref{sec:eb_joint_est} simplifies the selection of step sizes by providing their relative scalings, which allows the user to specify only a single global step size. 
The constraint sets in Algorithm~\ref{alg:joint_est} are defined as 
\begin{equation}\label{eq:joint_constraint}
\begin{aligned}
\mathcal{C}(\F):= \Big\{\F\in\mathbb{R}^{n\times r}\,|\, \|\F\|_{2,\infty}^2\leq M_0\Big\},&\quad 
\mathcal{C}(\Ld):= \Big\{\Ld\in\mathbb{R}^{p\times s}\,|\, \|\Ld\|_{2,\infty}^2\leq M_0,\ \Ld^\top 1_p  = 0_s\Big\},\\
\mathcal{C}(\L_m):= \Big\{\L_m\in\mathbb{R}^{r\times K_m}\,|\, \|\L_m\|_{\opn}^2\leq \frac{M_m}{2M_0}\Big\},&\quad \mathcal{C}(\R_m):= \Big\{\R_m\in\mathbb{R}^{s\times K_m}\,|\, \|\R_m\|_{\opn}^2\leq \frac{M_m}{2M_0}\Big\},\\
\mathcal{C}(\ap_m):= \Big\{\ap_m\in\mathbb{R}^{n}\,|\, \|\ap_m\|_{\infty}\leq \frac{M_m}{2}\Big\},&\quad\text{for $m=1,2$,}
\end{aligned}
\end{equation}
where $M_0,M_1,M_2>0$ are constants specific to the parameter space being considered (Section~\ref{sec:theory}). 
The projections onto $\mathcal{C}(\F)$, $\mathcal{C}(\L_m)$, $\mathcal{C}(\R_m)$ and $\mathcal{C}(\ap_m)$ admit closed-form solutions and can be computed efficiently. The projection onto $\mathcal{C}(\Ld)$, which involves two constraints, can be done using, e.g., Dykstra's alternating projection algorithm \citep{dykstra1983algorithm}. The projection step is essential for establishing theoretical guarantees. In practice, one can set $M_m$'s sufficiently large, or simply center $\Ld$ column-wise at the end of each iteration. We observe that dropping the other projections does not impair our algorithm's  performance.

\subsection{Initialization: Separate Estimation}\label{sec:sep_est}

Note that the combined loss \eqref{eq:combined_loss} is a non-convex function of $\Xi$. To obtain a good initialization for the non-convex joint estimation, we first separately estimate the two individual low-rank models~\eqref{eq:sep_model1} and \eqref{eq:sep_model2} on $Y$ and $Z$ respectively, treating $\F_1$, $\F_2$, $\Ld_1$, $\Ld_2$, $\alpha_1$ and $\alpha_2$ as independent parameters. We then construct $\Xi^0$ based on the resulting estimates $\{\hat{\F}_m, \hat{\Ld}_m, \hat{\ap}_m\}_{m=1,2}$. 
Specifically, let $U^{(r)}\Sigma^{(r)} (V^{(r)})^\top$ be the rank-$r$ truncated SVD of $(\hat{\F}_1\ \hat{\F}_2)\in\mathbb{R}^{n\times(K_1 + K_2)}$ so that $\Sigma^{(r)}\in \mathbb{R}^{r\times r}$. We initialize $\F$, $\L_1$ and $\L_2$ with $\F^0 = \mathcal{P}_{\mathcal{C}(\F)}(\sqrt{n}U^{(r)})$ and $\L_m^0 = \mathcal{P}_{\mathcal{C}(\L_m)}(\tilde{\L}_m^0)\ (m=1,2)$ where $(\tilde{\L}_1^0\ \tilde{\L}_2^0) := \Sigma^{(r)} (V^{(r)})^\top /\sqrt{n}$. $\Ld$, $\R_1$ and $\R_2$ are initialized similarly, and $\ap^0_m = \hat{\ap}_m$ for $m=1,2$.

Projected gradient descent has been employed for various low-rank matrix estimation problems \citep{chen2015fast, tu2016low, wang2017unified, chi2019nonconvex}, including the estimation of models \eqref{eq:sep_model1} and \eqref{eq:sep_model2}. 
Denoting $\Tt_m =\F_m\Ld_m^\top + \ap_m 1_p^\top$ for $m=1,2$, we perform projected gradient descent to minimize the penalized loss function
\begin{align*}
{\mathcal{L}}_m(\F_m, \Ld_m, \alpha_m) := L_m(\Tt_m) + \gamma np \Big\|\frac{\F_m^\top \F_m}{n} - \frac{\Ld_m^\top\Ld_m}{p}\Big\|_{\fn}^2,\quad m=1,2.
\end{align*}
Here $\gamma > 0$ is a tuning parameter, and $L_1(\Tt_1)$ and $L_2(\Tt_2)$ are loss functions defined in \eqref{eq:sep_loss1} and \eqref{eq:sep_loss2}, associated with the partially observed data matrix $Y$ and its missingness pattern $Z$, respectively. 
The additional penalty term is designed to force the estimates of $\F_m$ and $\Ld_m$ to be balanced instead of regularizing them, following \citet{tu2016low} and \citet{park2018finding}. Note that Condition~\ref{ass:ident_prod} implies $\F_m^\top \F_m/n = \Ld_m^\top \Ld_m/p$.
The tuning parameter $\gamma$ admits a wide range of viable values, as shown in Proposition~\ref{thm:sep_est}. All algorithmic details, including the initialization of the separate estimation are summarized in the appendix. 

\section{Theoretical Results}\label{sec:theory}

In this section, we present error bounds for both the joint estimation and its initialization under the joint low-rank model~\eqref{eq:model}. Specifically, we establish the local convergence properties of projected gradient descent for both the joint and separate estimation, under the following assumption on the data generating process. 

\begin{assumption}[Data generating process]\label{ass:DGP}
The data matrix $Y$ and its missingness pattern $Z$ are generated by model~\eqref{eq:model} with true parameter $\Xi_*\in \mathcal{F}(\mathcal{M})$ satisfying Condition~\ref{ass:ident}. The noise $E$ is independent of $Z$, and has independent sub-Gaussian entries with mean zero and variance proxy $b^2=O(1)$. 
\end{assumption}

The parameter space $\mathcal{F}(\mathcal{M})$, defined as 
\begin{equation}\label{eq:model_space}
\begin{aligned}
\mathcal{F}(\mathcal{M}) := \Big\{\Xi\,\Big|\,
&\|\F\|_{2,\infty}^2,\|\Ld\|_{2,\infty}^2\leq M_0;\ \|\L_m\|_{\opn}^2,\, \|\R_m\|_{\opn}^2\leq \frac{M_m}{2M_0}\,(m=1,2);\\
&\|\ap_m\|_{\infty}\leq \frac{M_m}{2}\,(m=1,2);\ \max_{i,j}\Tt_{2,ij}\leq - M_3\Big\},
\end{aligned}
\end{equation}
is determined by a set of quantities $\mathcal{M} = \{M_0, M_1, M_2, M_3\}$ in which $M_0, M_1, M_2>0$. In our theoretical analysis, we assume that $M_0$, $M_1$ and $M_2$ are known and the constraint sets in \eqref{eq:joint_constraint} are defined accordingly. The assumption on $\Xi_*$ is well-defined despite the identifiability issue of the model, because if $\Xi_*$ and $\tilde{\Xi}_*$ specify the same data distribution and they both satisfy Condition~\ref{ass:ident}, then by Proposition~\ref{prop:ident}, $\Xi_*\in\mathcal{F}(\mathcal{M})$ implies $\tilde{\Xi}_*\in\mathcal{F}(\mathcal{M})$. 
Note that $\Xi_*\in\mathcal{F}(\mathcal{M})$ implies 
$$
|\Tt_{1*, ij}|\leq M_1,\quad \text{and}\quad 
\frac{1}{2}e^{-M_2}\leq \mathbb{E}[Z_{ij}] = \sigma(\Tt_{2*,ij})\leq e^{-M_3}\wedge 1.
$$
We further define $\pi_{-} := e^{-M_2}/2$ and $\pi_{+} := e^{-M_3}\wedge 1$, representing the minimum and maximum probabilities of an entry in $Y$ being observed. $\pi_-$ characterizes the worst-case hardness of the completion, as roughly $\pi_-np$ entries of $Y$ will be observed at worst. The condition number $\pi_+/\pi_-$ represents the heterogeneity of the missingness. In terms of $Z$, $\pi_-$ and $\pi_+$ characterize the sparsity of the binary matrix.

\subsection{Error Bound for Joint Estimation}\label{sec:eb_joint_est}

Due to the identifiability issue described in Section~\ref{sec:ident}, we need to carefully define an error metric to eliminate the impact of the indeterminacy. Define ${\sigma}_{m+}:= \sigma_{+}(\F_{m*})/\sqrt{n} = \sigma_{+}(\Ld_{m*})/\sqrt{p}$ and $\sigma_{m-} := \sigma_{-}(\F_{m*})/\sqrt{n} = \sigma_{-}(\Ld_{m*})/\sqrt{p}$.
Since $\Xi_*$ satisfies Condition~\ref{ass:ident_prod}, we have ${\sigma}_{m+} = \sigma_{-}(\L_{m*}) = \sigma_{-}(\R_{m*})$ and $\sigma_{m-} = \sigma_{-}(\L_{m*}) = \sigma_{-}(\R_{m*})$ for $m=1,2$. We further denote  ${\sigma}_{+} := {\sigma}_{1+}\vee {\sigma}_{2+}$ and $\sigma_{-} := \sigma_{1-}\wedge \sigma_{2-}$. To characterize the evolution of iterates, we consider the following error metric,
\begin{equation}\label{eq:error_joint}
\begin{aligned}
e^2(\Xi):=&
\min_{\substack{O_{\F}\in\mathbb{O}^r, O_\Ld\in\mathbb{O}^s,\\
O_1\in\mathbb{O}^{K_1}, O_2\in\mathbb{O}^{K_2}}} p\sigma_+^4\|\F-\F_* O_{\F}\|_{\fn}^2 + n\sigma_+^4 \|\Ld-\Ld_* O_\Ld\|_{\fn}^2 + np\sigma_{1+}^2 \|\L_1 - O_{\F}^\top \L_{1*} O_1\|_{\fn}^2\\
&+ np\sigma_{1+}^2 \|\R_1 - O_\Ld^\top \R_{1*} O_1\|_{\fn}^2+ np\sigma_{2+}^2 \|\L_2 - O_\F^\top \L_{2*} O_2\|_{\fn}^2+ np\sigma_{2+}^2 \|\R_2 - O_\Ld^\top \R_{2*} O_2\|_{\fn}^2\\
&+ p\|\ap_1 - \ap_{1*}\|_2^2 + p\|\ap_2 - \ap_{2*}\|_2^2.
\end{aligned}
\end{equation}
The error $e^2(\Xi)$ is a weighted sum of the estimation errors in the (squared) Frobenius norm of all parameters, minimized over all equivalent parametrizations of the true model, restricted to those satisfying Condition~\ref{ass:ident} (see Proposition~\ref{prop:ident}).

Define $\Dt_{\Tt_m} = \Tt_m - \Tt_{m*}\ (m=1,2)$ and condition numbers
$$
\kappa_+ = {\sigma_{+}}/{\sigma_{-}}, \ \ 
\kappa_{\L} = \frac{\sigma_{+}((\L_{1*}\ \L_{2*}))}{\sigma_{-}((\L_{1*}\ \L_{2*}))},\ \ \kappa_{\R} = \frac{\sigma_{+}((\R_{1*}\ \R_{2*}))}{\sigma_{-}((\R_{1*}\ \R_{2*}))},\ \ 
{\kappa} = \kappa_\L\vee \kappa_\R. 
$$
Note that the order of $\kappa$ can be larger than that of $\kappa_+$, as $\sigma_{+}\leq \sigma_{+}((\L_{1*}\ \L_{2*})), \sigma_{+}((\R_{1*}\ \R_{2*}))\leq \sqrt{2}\sigma_{+}$, while $\sigma_{-}((\L_{1*}\ \L_{2*}))$ and $\sigma_{-}((\R_{1*}\ \R_{2*}))$, compared to $\sigma_{1-}\wedge \sigma_{2-}$, further depend on the principal angles between the column spaces of $\L_{1*}$ and $\L_{2*}$, and those of $\R_{1*}$ and $\R_{2*}$, respectively. 
The rationale behind the use of the proposed error metric and the penalty term $P(\Xi)$ defined in Section~\ref{sec:joint_est} can be partly seen from the following lemma.

\begin{lemma}[Equivalence of error metrics]\label{lemma:equiv_metric_main}

Suppose that $\Xi_*$ satisfies Condition~\ref{ass:ident}. Let $\Xi$ be an arbitrary set of parameters with $\Ld$ column-wise centered, i.e., $\Ld^\top 1_p = 0_s$. We have:
\begin{enumerate}
\item[(i)] if $\sigma_+(\L_m)\vee \sigma_+(\R_m) \leq (1+\delta)\sigma_{m+}\ (m=1,2)$ for some constant $\delta>0$, then there exist some constants $c_0>0$ and $\gamma_0>0$ such that 
\begin{align*}
&\|\Dt_{\Tt_1}\|_{\fn}^2 + \|\Dt_{\Tt_2}\|_{\fn}^2 + \gamma_0 np P_*(\Xi)\geq c_0\kappa^{-2}\kappa_+^{-2}e^2(\Xi),
\end{align*}
where
$$
P_*(\Xi) = \sigma_+^4 \Big\|\frac{\F^\top \F}{n}-I_r\Big\|_{\fn}^2 
+ \sigma_+^4\Big\|\frac{\Ld^\top \Ld}{p}-I_s\Big\|_{\fn}^2
+ \|\L_1^\top \L_1-\R_1^\top \R_1\|_{\fn}^2 
+ \|\L_2^\top \L_2-\R_2^\top \R_2\|_{\fn}^2;
$$
\item[(ii)] if in addition $\sigma_+(\F)\leq (1+\delta)\sqrt{n}$ and $\sigma_+(\Ld) \leq (1+\delta)\sqrt{p}$, then there exists a constant $C_0>0$, such that
\begin{align*}
\|\Dt_{\Tt_1}\|_{\fn}^2 + \|\Dt_{\Tt_2}\|_{\fn}^2 \leq C_0 e^2(\Xi).
\end{align*}
\end{enumerate}
\end{lemma}

Lemma~\ref{lemma:equiv_metric_main} states that with the penalty term $P_*(\Xi)$, controlling $e^2(\Xi)$ is equivalent to controlling the estimation error in the lifted space, $\|\Dt_{\Tt_1}\|_{\fn}^2 + \|\Dt_{\Tt_2}\|_{\fn}^2$. Specifically, $e^2(\Xi)$ measures the estimation error of the factorized parameters $\Xi$ updated by the algorithm. Leveraging the strong convexity of $L_1(\Tt_1)$ and $L_2(\Tt_2)$ as functions of $\Tt_1$ and $\Tt_2$, we can show that at each iteration, $e^2(\Xi)$ will decrease by an amount proportional to $\|\Dt_{\Tt_1}\|_{\fn}^2 + \|\Dt_{\Tt_2}\|_{\fn}^2$. Lemma~\ref{lemma:equiv_metric_main} and the penalty term $P_*(\Xi)$ then allow us to link the amount of descent back to $e^2(\Xi)$, which eventually leads to the contraction of $e^2(\Xi)$ (Theorem~\ref{thm:joint_est}). 
The above argument motivates us to use the penalty term $P_*(\Xi)$. However, $\sigma_{+}$ in $P_*(\Xi)$ depends on the unknown true parameters. With a good initialization $\Xi^0$, a natural solution is to replace $\sigma_{+}$ with an estimate, $\hat{\sigma}_+$, where $\hat{\sigma}_{+} = \hat{\sigma}_{1+}\vee \hat{\sigma}_{2+}$, and $\hat{\sigma}_{1+}:= \sigma_+(\L_1^0)\vee \sigma_+(\R_1^0)$, $\hat{\sigma}_{2+}:= \sigma_+(\L_2^0)\vee \sigma_+(\R_2^0)$. 
Therefore, we consider setting $\gamma_1 = \gamma \hat{\sigma}_+^4$ and $\gamma_2 = \gamma$ in $P(\Xi)$ with some constant $\gamma > 0$. 
As shown in Theorem~\ref{thm:joint_est}, $\gamma$ is viable across a wide range of values, and in our simulation we also find that the performance of the algorithm is insensitive to its choice. In practice, one may simply set $\gamma = 0.1$. 

We establish the theoretical guarantee for Algorithm~\ref{alg:joint_est} with the step sizes set as
\begin{equation*}
\begin{aligned}
\eta_{\F} &= \frac{\eta}{p\hat{\sigma}_+^{4}},\ \ \eta_\Ld = \frac{\eta}{n\hat{\sigma}_+^{4}},\ \ \eta_{\L_1} = \frac{\eta}{np\hat{\sigma}_{1+}^{2}},\ \ \eta_{\R_1} = \frac{\eta}{np\hat{\sigma}_{1+}^{2}},\\
\eta_{\L_2} &= \frac{\eta}{np\hat{\sigma}_{2+}^{2}},\ \ \eta_{\R_2} = \frac{\eta}{np\hat{\sigma}_{2+}^{2}},\ \  \eta_{\ap_1} = \frac{\eta}{p},\ \ \eta_{\ap_2} = \frac{\eta}{p}, 
\end{aligned}
\end{equation*}
where $\eta>0$ is some small constant. Similarly, here $\hat{\sigma}_+$, $\hat{\sigma}_{1+}$ and $\hat{\sigma}_{2+}$ are estimates of the corresponding unknown true values. In our numerical study we find that $\eta = 0.1$ generally works well. 
Our theoretical result relies on the following assumption that the parameters are properly initialized. 

\begin{assumption}[Sufficiently good initialization]\label{ass:init_joint}
Denote $w_*:= \min\{w,1-w\}$. The initialization $\Xi^0$ is good enough so that for some small constant $\delta>0$,
\begin{align}\label{eq:init_joint}
\frac{e^2(\Xi^0)}{np}\leq \delta \kappa^{-2}\kappa_+^{-2}w_*\pi_-(\sigma_{1+}^4 \wedge \sigma_{2+}^4).
\end{align}
Furthermore, the parameters in $\Xi^0$ belong to the constraint sets defined in \eqref{eq:joint_constraint}. 

\end{assumption}

\begin{theorem}\label{thm:joint_est} 

Suppose that Assumptions~\ref{ass:DGP} and \ref{ass:init_joint} hold, $\pi_+ = \Omega\big((n\vee p)^{-1}{\log^2(n\vee p)}\big)$ and $M_1 = O(1)$. Let $w\in(0,1)$ and $w_*:= \min\{w,1-w\}$. Choose constants $\gamma > c_1 w_*\pi_-$ for some constant $c_1>0$ and $\eta>0$ sufficiently small. Then there exists a constant $C_1>0$ such that, if
\begin{align}\label{eq:LB_sample_size_joint}
n\wedge p\geq \frac{C_1\kappa^4 \kappa_+^4\pi_+}{(\sigma_{1+}^4 \wedge \sigma_{2+}^4)\pi_-}\frac{K_1\vee K_2}{w_*^2\pi_-^2}, 
\end{align}
with probability at least $1-10(n+p)^{-2}$, we have
\begin{align*}
e^2(\Xi^{t}) \leq 3\Big(1 - c \eta \kappa^{-2}\kappa_{+}^{-2}w_*\pi_- \Big)^t
e^2(\Xi^{0}) + C \kappa^{2}\kappa_{+}^{2} \frac{\pi_+}{w_*\pi_-^2}(K_1\vee K_2)(n\vee p),
\end{align*}
for some constants $c, C>0$ and any $t\in\mathbb{Z}_+$. 

\end{theorem}

Theorem~\ref{thm:joint_est} states that the estimation error of the iterates consists of two parts: an algorithmic error converging linearly to zero and a statistical error introduced by random noise. 
Treating $\sigma_{1+}$, $\sigma_{2+}$, $\kappa$, $\kappa_+$ and $w_*$ as constants, the theorem implies that if the number of iterations $T$ is of order
$\Omega\big(\pi_-^{-1} \log\big[ \pi_-^3 \pi_+^{-1}(K_1\vee K_2)^{-1} (n\wedge p) \big]\big)$, we have
\begin{align}\label{eq:MSE_joint}
\frac{1}{np}\Big(\|\Dt_{\hat{\Tt}_1}\|_{\fn}^2 + \|\Dt_{\hat{\Tt}_2}\|_{\fn}^2\Big)
\lesssim \frac{e^2(\hat{\Xi})}{np} = 
O_P\Big(\frac{\pi_+}{\pi_-^2} \frac{K_1\vee K_2}{n\wedge p}\Big),
\end{align} 
where the first inequality follows from Lemma~\ref{lemma:equiv_metric_main}. For $\Tt_1$, the rate in \eqref{eq:MSE_joint} matches (with $K_1\vee K_2$ replaced by $K_1$) the upper bound for matrix completion in \citet{klopp2015}, which is based on nuclear norm regularization. They assume that $\Tt_{1*}\in\{\Tt\in\mathbb{R}^{n\times p}\,|\, \rank(\Tt)\leq K_1,\ \|\Tt\|_{\infty}\leq M_1\}$ and $M_1 = O(1)$, which is implied by our assumption on the parameter space. Their bound is minimax optimal when the missingness is  asymptotically homogeneous, i.e., $\pi_+/\pi_- = \Theta(1)$ \citep{Koltchinskii2011}. For $\Tt_2$, the rate in \eqref{eq:MSE_joint} is aligned with (with $K_1\vee K_2$ replaced by $K_2$) the upper bound for non-linear low-rank matrix estimation established by \citet{chen2022determining}, which studies the non-convex maximum likelihood estimator. By adapting the proof of Proposition 1 therein, it can be shown that their result is minimax optimal when the sparsity pattern is asymptotically homogeneous, i.e., $\pi_+/\pi_- = \Theta(1)$. Regarding $\Tt_2$, the parameter space we consider here is slightly more restricted than theirs due to the additional shared linear structure in our model.

An important observation from the bound in \eqref{eq:MSE_joint} and the above argument on its optimality is that the estimation errors of $\Tt_1$ and $\Tt_2$ have the same dependency on $\pi_-$ and $\pi_+$. 
Intuitively, this implies that the missingness in $Y$ (or equivalently, the sparsity in $Z$) has the same negative impact on the estimation of $\Tt_1$ and $\Tt_2$, a property stemming from the logistic link function. 
Due to the similar roles that missingness and sparsity play in reducing the sample size, the estimation error rate of our joint model is not expected to improve upon existing results for the two individual models \citep{klopp2015, chen2022determining}, as the number of parameters is only reduced by a constant fraction by incorporating the shared structure, while the effective sample sizes from both sub-models are of the same order.
For the same reason, while our result holds for any arbitrary fixed weight $w\in(0,1)$, the optimal weight that yields the fastest algorithmic convergence and minimum statistical error (in terms of the rate) is of constant order.

Note that the error bound in \eqref{eq:MSE_joint} depends linearly on the condition number $\pi_+/\pi_-$. In principle, this dependency can be removed through inverse probability weighting (IPW) \citep{mao2021matrix, wang2021matrix}. Specifically, if the true observation probabilities $(\pi_{ij})_{i\in[n],j\in[p]} := \mathbb{E}[Z]$ were known, the target matrix $\Tt_{1*} = \mathbb{E}[Y]$ could be estimated by minimizing the following debiased least-squares loss with proper regularization,
\begin{align*}
L_{\text{IPW}}(\Tt_1) := \frac{1}{2}\sum_{i,j}{\pi}_{ij}^{-1}Z_{ij}(Y_{ij} - \Tt_{1,ij})^2.
\end{align*}
While \citet{mao2021matrix} show that the error bound of such an oracle IPW estimator no longer depends on $\pi_+/\pi_-$, in practice, the $\pi_{ij}$'s need to be estimated. They assume a non-linear low-rank model on $Z$ and estimate its parameter matrix through nuclear norm regularization. 
However, since the $\pi_{ij}$'s are typically close to zero, the resulting IPW estimator suffers from high variance. 
To mitigate this issue, \citet{mao2021matrix} propose to re-estimate the low-rank model on $Z$ under a tighter constraint, which controls the extreme probabilities at the price of misspecifying the model. Nonetheless, determining the optimal constraint that balances this trade-off is tricky, and the error bound of their IPW estimator with estimated $\hat{\pi}_{ij}$'s still retains a linear dependency on $\pi_+/\pi_-$.

For the error bound in \eqref{eq:MSE_joint} to be valid, there is an additional  requirement~\eqref{eq:LB_sample_size_joint} on the sample size. 
Assuming that $K_1$, $K_2$ and $\pi_+/\pi_-$ are constants, it simplifies to $\pi_- = \Omega((n\wedge p)^{-1/2})$, which is more stringent than the order $\pi_- = \Omega((n\wedge p)^{-1})$ needed to ensure that the upper bound in \eqref{eq:MSE_joint} vanishes. 
This stronger requirement is likely an artifact of the algorithmic analysis, as it is not needed for the consistency of the global minimizer (see Theorem~\ref{thm:IC_joint}).
A similar requirement~\eqref{eq:LB_sample_size_sep} also appears in the theoretical guarantee for separate estimation (see Proposition~\ref{thm:sep_est}). 
We defer the detailed discussion of these requirements to the appendix.

Finally, Assumption~\ref{ass:init_joint} requires a good initialization satisfying $(np)^{-1} e^2(\Xi^0) = O(\pi_-)$. When the sample size requirement \eqref{eq:LB_sample_size_joint} is satisfied, the initialization error $O(\pi_-)$ is no smaller than the error bound in \eqref{eq:MSE_joint}
for the estimator at convergence. 
Therefore, Algorithm~\ref{alg:joint_est} only requires an initialization with a slower rate of consistency. In particular, if $\pi_-=\Theta(1)$, i.e., a constant fraction of entries in $Y$ will always be observed, then any initialization with statistical consistency suffices, given a sufficiently large sample size.

\subsection{Error Bound for Initialization}\label{sec:eb_sep}

In this subsection, we present the local convergence guarantee for the separate estimation, which shows that the initialization for joint estimation satisfies Assumption~\ref{ass:init_joint} and is thus valid under certain conditions. 
To characterize the convergence of projected gradient descent, we consider the following error metric,
$$
e^2(\F_m, \Ld_m, \ap_m) := \sigma_{m+}^2 \min_{O\in\mathbb{O}^{K_m}}\{p\|\F_m - \F_{m*}O\|_{\fn}^2 + n \|\Ld_m - \Ld_{m*}O\|_{\fn}^2\} + p\|\ap_m - \ap_{m*}\|_2^2.
$$
The rationale here is similar to that for $e^2(\Xi)$ in the joint estimation.
Define the condition numbers $\kappa_{m} = {\sigma_{m+}}/{\sigma_{m-}}$ for $m=1,2$. 
We assume the algorithm is properly initialized.

\begin{assumption}[Sufficiently good initialization]\label{ass:init_sep}
The initialization $\{\F^0_m, \Ld^0_m, \ap^0_m\}_{m=1,2}$ is good enough so that for some small constant $\delta>0$, 
\begin{align}\label{eq:init_sep}
\frac{e^2(\F_m^0, \Ld_m^0, \ap_m^0)}{np}\leq \delta \kappa^{-2}_m\pi_{-}\sigma_{m+}^4.
\end{align}
Furthermore, $\F^0_m$, $\Ld^0_m$ and $\ap^0_m$ belong to the constraint sets defined in the appendix. 

\end{assumption}

The following proposition shows that projected gradient descent achieves a linear convergence rate to a statistically good solution, given a proper initialization and a sufficiently large sample size. It follows from the general framework for analyzing low-rank matrix estimation developed by \citet{wang2017unified} and \citet{park2018finding}.

\begin{proposition}\label{thm:sep_est}
Suppose that Assumptions~\ref{ass:DGP} and \ref{ass:init_sep} hold, $\pi_+ = \Omega\big((n\vee p)^{-1}{\log^2(n\vee p)}\big)$ and $M_1 = O(1)$. Choose constant $\gamma > c_1 \pi_-$ for some constant $c_1 > 0$. 
Let $\{\F_m^t, \Ld_m^t, \ap_m^t\}$ be the $t$-th iteration of the projected gradient descent algorithm described in the appendix. 
Then there exists a constant $C_1>0$ such that, if
\begin{align}\label{eq:LB_sample_size_sep}
n\wedge p \geq \frac{C_1\kappa_m^4\pi_+}{\sigma_{m+}^4\pi_-}\frac{K_m}{\pi_-^2},
\end{align}
with probability at least $1-10(n+p)^{-2}$, we have 
\begin{align*}
e^2(\F_m^t, \Ld_m^t, \ap_m^t)\leq 3\Big( 1 - c\eta\kappa_{m}^{-2}\pi_- \Big)^t e^2(\F_m^0, \Ld_m^0, \ap_m^0)
+ C \kappa_m^2 K_m\frac{\pi_+}{\pi_-^2} (n\vee p),
\end{align*}
for $m=1,2$, some constants $c,C>0$ and any $t\in\mathbb{Z}_+$. Here $\eta>0$ is a step size constant.

\end{proposition}

Treating $\sigma_{m+}$ and $\kappa_m\ (m=1,2)$ as constants, the proposition implies that with a sufficiently large number of iterations $T$, we have 
\begin{align}\label{eq:MSE_sep}
\frac{\|\hat{\Tt}_m - \Tt_{m*}\|_{\fn}^2}{np}\lesssim 
\frac{e^2(\hF_m, \hLd_m, \hap_m)}{np}
= O_P\Big(\frac{\pi_+}{\pi_-^2} \frac{K_m}{n\wedge p}\Big), \quad m=1,2,
\end{align}
where $\{\hat{\F}_m, \hat{\Ld}_m, \hat{\ap}_m\} := \{\F^T_m, \Ld^T_m, \ap^T_m\}$ is the output of the algorithm. 
This error rate is identical to that of the joint estimation in \eqref{eq:MSE_joint}, with $K_1\vee K_2$ replaced by $K_m$. Therefore, most earlier discussion regarding this rate remains applicable. Recall that the initialization of joint estimation $\Xi^0$ is obtained by transforming $\{\hF_m, \hLd_m, \hap_m\}_{m=1,2}$. It is shown in the appendix 
that
\begin{align}\label{eq:bound_trans}
e^2(\Xi^0)\leq \frac{32 \kappa^2 \sigma_+^2}{\sigma_{1+}^2\wedge \sigma_{2+}^2} \Big[ e^2(\hF_1, \hLd_1, \hap_1) + e^2(\hF_2, \hLd_2, \hap_2)\Big],  
\end{align}
so the transformation only scales the estimation error by a constant factor. Consequently, we have $(np)^{-1}e^2(\Xi^0) = O_P\big({\pi_+}{\pi_-^{-2}} (n\wedge p)^{-1}{(K_1\vee K_2)}\big) = O_P(\pi_-)$, indicating that the proposed $\Xi^0$ is a valid initialization with high probability.

Similar to the joint estimation, the separate estimation requires a good initialization with error $(np)^{-1}e^2(\F_m^0, \Ld_m^0, \ap_m^0)$ of order $O(\pi_-)$ for $m=1,2$, a rate no faster than $O\big({\pi_+}{\pi_-^{-2}} {(n\wedge p)^{-1}}{K_m}\big)$ under the stated assumptions. We present possible initialization methods in the appendix. 

\section{Selection of Latent Dimensions}\label{sec:model_selection}

In our estimation approach, the latent dimensions $K_1$, $K_2$, $r$ and $s$ need to be properly determined. In practice, they can be selected via cross-validation, which splits the entries of $Y$ and $Z$ into training and validation sets. In order to avoid the computational burden of cross-validation, we propose a criterion-based two-step procedure for model selection, which first estimates $K_1$ and $K_2$ treating the two models separately, then estimates $r$ and $s$ in the joint model. A notable advantage of the proposed procedure is that $K_1$, $K_2$, $r$ and $s$ can be determined sequentially, so no grid search needs to be performed. We describe the procedure in detail first, and then present its theoretical guarantee. 

\subsection{Criterion-Based Model Selection}

We first consider the two individual low-rank models. For any alternative $K_m\ (m=1,2)$ satisfying $1\leq K_m\leq K_{m+}$, we denote the output of separate estimation by $\hTt_m^{(K_m)} = \hF_m^{(K_m)}(\hLd_m^{(K_m)})^\top + \hap_m^{(K_m)}1_p^\top$. 
Here $K_{m+}$ is some pre-specified upper bound, and we allow $K_{m+} = +\infty$. We determine the optimal $K_m$ by minimizing the following criterion,
\begin{align}\label{eq:IC_sep}
\hat{K}_m := \argmin_{1\leq K_m\leq K_{m+}} L_m(\hTt_m^{(K_m)}) + \rho K_m(n\vee p) \log\left(\frac{\sum_{i,j}Z_{ij}}{n\vee p}\right),
\end{align}
in which $\rho>0$ is some pre-specified constant. This is essentially the Joint-likelihood-based Information Criterion (JIC) proposed by \citet{chen2022determining} for determining the number of factors in generalized latent factor models in the presence of missing data.

We then estimate $r$ and $s$ in the joint model, fixing $K_1 = \hat{K}_1$ and $K_2=\hat{K}_2$. Recall the constraint $K_1\vee K_2\leq r,s \leq K_1 + K_2$. To estimate $r$, we over-specify $s = K_1 + K_2$ and fit the joint model using Algorithm~\ref{alg:joint_est}. Denoting the output by $\hTt_m^{(r)} = \hF^{(r)} \hL_m^{(r)}(\hR_m^{(r)})^\top(\hLd^{(r)})^\top + \hap_m^{(r)}1_p^\top$, we determine the optimal $r$ by minimizing the following criterion,
\begin{align}\label{eq:IC_joint}
\hat{r} := \argmin_{K_1\vee K_2\leq r\leq K_1 + K_2} w L_1(\hTt_1^{(r)}) + (1-w) L_2(\hTt_2^{(r)}) + 
\tilde{\rho}\, r(n\vee p) \log\left(\frac{\sum_{i,j}Z_{ij}}{n\vee p}\right),
\end{align}
where $\tilde{\rho}>0$ is some pre-specified constant that may differ from $\rho$. 
The latent dimension $s$ can be estimated in the same manner by over-specifying $r=K_1 + K_2$ and performing the same procedure with the roles of $r$ and $s$ flipped. 
We note that although theoretically the specific choices of $\rho$ and $\tilde{\rho}$ are inconsequential given a sufficiently large sample size (see Theorem~\ref{thm:IC_joint}), in finite samples they still need to be determined with care. In practice, we suggest setting $\rho=0.5$, as in \citet{chen2022determining}, and $\tilde{\rho}=0.1$.

\subsection{Theoretical Guarantee}

In this subsection, we assume that the data generation satisfies Assumption~\ref{ass:DGP} with true latent dimensions $K_{1*}$, $K_{2*}$, $r_{*}$, and $s_{*}$. 
Here we no longer require the true parameters to satisfy Conditions~\ref{ass:ident_cen} and \ref{ass:ident_prod}, as our main focus is the asymptotic behavior of the empirical loss values.
The selection consistency for $K_1$ and $K_2$ is a direct consequence of the results in \citet{chen2022determining},
so we only present the theoretical guarantee for the selection of $r$. In the following, we assume that $K_{1*}$ and $K_{2*}$ are known and no longer distinguish between $K_m$ and $K_{m*}\ (m=1,2)$. 

Fix some $s_* \leq s \leq K_1 + K_2$ throughout this subsection. 
For simplicity, we denote $L_w(\Tt_1, \Tt_2) := w L_1(\Tt_1) + (1-w) L_2(\Tt_2)$. 
Let $\hTt_m^{(r)} = \hF^{(r)} \hL_m^{(r)} (\hR_m^{(r)})^\top (\hLd^{(r)})^\top + \hap_m^{(r)}1_p^\top\ (m=1,2)$ be an arbitrary estimate of $\Tt_{m*}$ with candidate latent dimension $r$ satisfying $K_1\vee K_2\leq r\leq K_1 + K_2$. We impose the following assumption on it. 

\begin{assumption}\label{ass:IC_joint}

For all $r_*\leq r\leq K_1 + K_2$, it holds that $L_w(\hTt_1^{(r)}, \hTt_2^{(r)}) \leq L_w(\Tt_{1*}, \Tt_{2*})$ and
\begin{align*}
\hF^{(r)}\in \mathcal{C}(\F),\ \hLd^{(r)}\in \tilde{\mathcal{C}}(\Ld),\  \hL^{(r)}_m\in\mathcal{C}(\L_m),\ \hR^{(r)}_m\in\mathcal{C}(\R_m), \ \hap_m^{(r)}\in \mathcal{C}(\ap_m), 
\end{align*}
for $m=1,2$. Here $\tilde{\mathcal{C}}(\Ld):= \{\Ld\in\mathbb{R}^{p\times s}\,|\, \|\Ld\|_{2,\infty}^2\leq M_0\}$ is obtained by removing the identifiability constraint $\Ld^\top 1_p =0_{s}$ from $\mathcal{C}(\Ld)$.
\end{assumption}

Assumption~\ref{ass:IC_joint} requires that, when $r$ is not under-specified, the estimated parameters lie in certain feasible sets, and $\hTt_m^{(r)}$'s have a smaller loss value than the true parameters. Clearly the minimizer of 
$L_w(\Tt_1, \Tt_2)$ under the constraints in Assumption~\ref{ass:IC_joint} satisfies this condition. Our Algorithm~\ref{alg:joint_est} aims to minimize $\mathcal{L}(\Xi)$, the sum of $L_w(\Tt_1, \Tt_2)$ and a penalty term, and by design its output satisfies the constraints in Assumption~\ref{ass:IC_joint}. Hence, as long as Algorithm~\ref{alg:joint_est} returns an estimate such that $\mathcal{L}(\hat{\Xi}^{(r)})\leq \mathcal{L}(\Xi_*)$, Assumption~\ref{ass:IC_joint} will be satisfied. However, we acknowledge that our convergence analysis of Algorithm~\ref{alg:joint_est} applies only to the case where the latent dimensions are correctly specified. In practice, we observe similar convergence behaviors regardless of whether they are over-specified or not, while rigorous theoretical justification remains an interesting future research problem.

The following theorem establishes the selection consistency of the proposed information criterion for $r$.

\begin{theorem}\label{thm:IC_joint}
Suppose that the data generation satisfies Assumption~\ref{ass:DGP}, while the true parameters need not satisfy Conditions~\ref{ass:ident_cen} and \ref{ass:ident_prod}. 
Assume that $\pi_+ = \Omega\big((n\vee p)^{-1}{\log^2(n\vee p)}\big)$ and $M_1 = O(1)$. Let $w\in(0,1)$ be any fixed constant. Then for any estimator of $\Tt_{m*}$ satisfying Assumption~\ref{ass:IC_joint}, there exist universal constants $C_1,C_2, C_3, C_4>0$ such that, with probability at least $1-2(n+p)^{-2}$, we have:
\begin{enumerate}
\item[(i)] for all $r_{*}\leq r\leq K_{1} + K_2$, 
$
\|\hTt_1^{(r)} - \Tt_{1*}\|_{\fn}^2 + \|\hTt_2^{(r)} - \Tt_{2*}\|_{\fn}^2\leq C_1{\pi_+}{\pi_{-}^{-2}}(r\wedge s)(n\vee p)
$,
and 
$$
L_w(\hTt_1^{(r)}, \hTt_2^{(r)}) - L_w(\hTt_1^{(r_*)}, \hTt_2^{(r_*)})
\geq -C_2{\pi_+}{\pi_{-}^{-1}}(r\wedge s)(n\vee p);
$$
\item[(ii)] if in addition $\sigma_{r_{*}+2}^2((\Tt_{1*}\ \Tt_{2*}))\geq C_3{\pi_+}{\pi_{-}^{-2}}(r_{*}\wedge s)(n\vee p)$, then for all $K_1\vee K_2\leq r <r_{*}$,
$$
L_w(\hTt_1^{(r)}, \hTt_2^{(r)}) - L_w(\hTt_1^{(r_*)}, \hTt_2^{(r_*)})
\geq C_4 \pi_- (r_{*} - r) \sigma_{r_{*}+2}^2((\Tt_{1*}\ \Tt_{2*})).
$$
\end{enumerate}
Consequently, consider the following estimator
\begin{align*}
\hat{r} := \argmin_{K_1\vee K_2\leq r\leq K_1+K_2} 
L_w(\hTt_1^{(r)}, \hTt_2^{(r)})
+ \tilde{\rho}\, r(n\vee p) h(n,p),
\end{align*}
where $\tilde{\rho}>0$ is any constant and $h(n,p)$ may be random. If in addition $\pi_+/\pi_-,r_*=\Theta(1)$, $h(n,p)\to_P +\infty$ and $h(n,p)=o_{P}\big(\pi_-\sigma_{r_* + 2}^2((\Tt_{1*}\ \Tt_{2*}))/(n\vee p)\big)$, then $\mathbb{P}(\hat{r} = r_{*})\to 1$. 
\end{theorem}

For any estimator satisfying Assumption~\ref{ass:IC_joint}, Theorem~\ref{thm:IC_joint} provides estimation error bounds for correctly and over-specified $r$, and characterizes the asymptotic behavior of the empirical loss value across the entire range of possible $r$. Selection consistency for a general information criterion follows as a direct consequence, provided that the asymptotic order of $h(n, p)$ is appropriately bounded. Specifically, under the standard assumption that $\sigma_{r_* + 2}^2((\Tt_{1*}\ \Tt_{2*})) = \Omega(np)$, a valid $h(n, p)$ exists as long as $\pi_- = \omega((n \wedge p)^{-1})$. Our proposed criterion adopts $h(n,p) = \log\big((n\vee p)^{-1}\sum_{i,j}Z_{ij}\big)$, which satisfies the requirement $h(n,p) \to_P +\infty$ at a sufficiently slow rate to ensure the upper bound condition is met under broad settings.

\section{Simulation Studies}\label{sec:simulation}

In this section, we present simulation results for our proposed joint estimation approach and model selection procedure. 
Throughout the section, we vary $n$, $K_1$ and $r$, while fixing $p = \lfloor 3n/4\rfloor$, $K_2 = \lfloor 4K_1/3\rfloor$ and $s = r$. Given fixed $K_1$, $r$ and a sequence of $n$, the true parameters $\Xi_*$ are generated according to the following steps. 
\begin{enumerate}
\item Generate the coefficients of linear transformations $\L_{1*} \in \mathbb{R}^{r \times K_1}$, $\R_{1*} \in \mathbb{R}^{s \times K_1}$, $\L_{2*} \in \mathbb{R}^{r \times K_2}$ and $\R_{2*} \in \mathbb{R}^{s \times K_2}$ as random orthogonal matrices. Specifically, for $\L_{1*}$, a temporary matrix of size $r \times K_1$ is first generated with entries drawn i.i.d. from $\mathrm{Unif}(-1, 1)$. Letting $U \Sigma V^T$ be its SVD, we set $\L_{1*} = U V^T$.  This procedure is repeated independently to generate one set of $\{ \L_{1*}, \R_{1*}, \L_{2*}, \R_{2*} \}$.

\item Generate the unified embeddings $\F_* \in \mathbb{R}^{n \times r}$ and $\Ld_* \in \mathbb{R}^{p \times s}$ with entries drawn i.i.d. following $\mathrm{Unif}(-1,1)$ for each $n$ (and its corresponding $p$) in the sequence. 

\item Generate the intercepts $\ap_{1*},\ap_{2*} \in \mathbb{R}^n$  with entries drawn i.i.d. from $\mathrm{Unif}(-0.5, 0.5)$ and  $\mathrm{Unif}(a - 0.5, a + 0.5)$ for some constant $a$, respectively, for each $n$ in the sequence. 

\item For $m=1,2$, define $\Tt_{m*} =  c_m^2\F_* \L_{m*} \left( \Ld_* \R_{m*} \right)^\top / \sqrt{K_m} + \ap_{m*} 1_p^\top$ for each $n$, with the same set of $\{ \L_{1*}, \R_{1*}, \L_{2*}, \R_{2*} \}$. Here $c_1 = 1.4$ and $c_2 = 1.2$. 
\end{enumerate}
The above procedure produces one \textit{instance} of true parameters, in which the same set of linear transformations $\{ \L_{1*}, \R_{1*}, \L_{2*}, \R_{2*} \}$ is shared across all values of $n$. We then simulate \textit{realizations} of the (complete) data matrix $Y \in \mathbb{R}^{n \times p}$ and its missingness pattern $Z \in \{0,1\}^{n \times p}$ for each $n$. Specifically, we sample $Y_{ij} \sim \mathcal{N}(\Tt_{1*,{ij}}, \sigma_{ij}^2)$ with $\sigma_{ij} \sim \mathrm{Unif}(0.75, 1.5)$ and $Z_{ij} \sim \mathrm{Bernoulli}\left(\sigma(\Tt_{2*,ij})\right)$, independently for all entries. The final observed data is formed by masking entries in $Y$ where $Z_{ij} =0$. 
In the generation of $Z$, the constant $a$ controls the overall sparsity level of the missingness pattern, i.e., the observation rate in the data matrix. In all simulations, we vary $a\in\{-1.9, -0.6\}$, which correspond to approximately 15\% and 35\% of entries being observed, respectively.

For both sub-models, we compute the mean squared error (MSE) to evaluate estimators of $\Tt_{1*}$ and $\Tt_{2*}$ 
(where applicable), defined as
\begin{align}\label{eq:simu_MSE}
\mathrm{MSE}(\hTt_{m}, \Tt_{m*}) := \frac{1}{np} \|\hTt_{m}- \Tt_{m*}\|_{\fn}^2,
\end{align}
for $m=1,2$. For each experimental setting, we report results averaged over $20$ data realizations for a single instance of the true parameters. Results based on different parameter instances are provided in the appendix. 
These results all exhibit consistent patterns, supporting our arguments in this section.

\subsection{Performance of Separate and Joint Estimation}

In this subsection, we numerically study the performance of the separate and joint estimation approaches, with a focus on their asymptotic behavior and the impact of the shared linear structure. We vary $n\in\{300, 600, 900, 1200, 1500\}$, and present the estimation errors of both the separate and joint estimations. Recall that the output of separate estimation is used as the initialization for joint estimation.

\begin{figure}[htb!]
\centering
\includegraphics[width=0.85\textwidth]{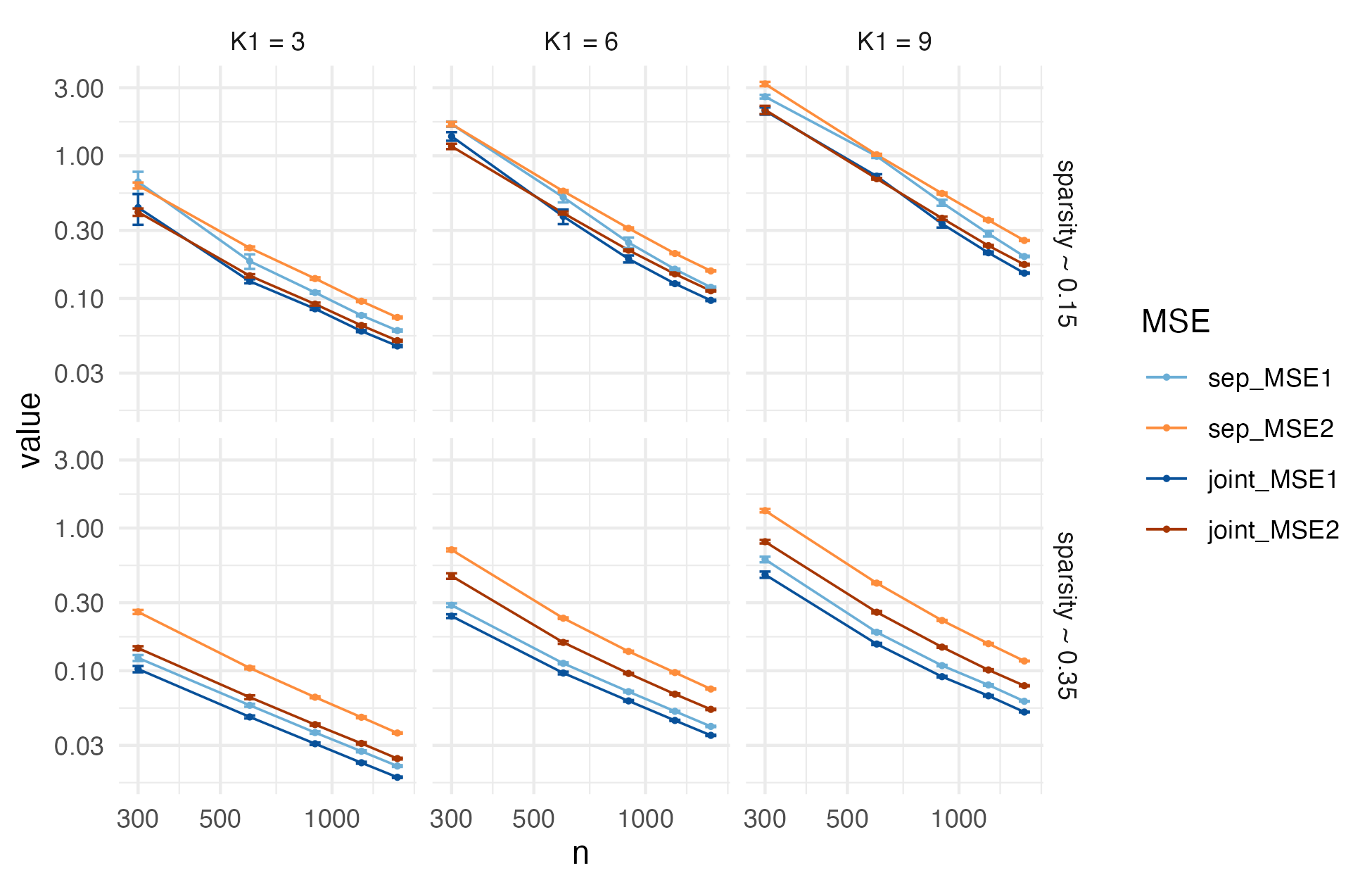}
\caption{Mean Squared Errors (MSEs) for $\Tt_{1*}$ and $\Tt_{2*}$ as a function of $n$. Panels show results for varying latent dimensions ($K_1$) and observation rates (sparsity), with median-sized $r$. Separate and joint estimations are represented by light- and dark-colored lines, respectively. Both axes are on a $\log_{10}$-scale. Error bars represent $\pm 1$ standard error over $20$ data realizations.}\label{fig:asym_fix_r} 
\end{figure}

In the first experiment, we vary the latent dimension $K_1\in\{3,6,9\}$. For each $K_1$, we set $r$ to the midpoint of its allowable range $K_1\vee K_2\leq r\leq K_1 + K_2$, yielding $r\in\{5,11,15\}$. This creates a median-sized shared linear structure. The results are presented in Figure~\ref{fig:asym_fix_r}. 
They are well aligned with our theoretical analysis. In all cases, the MSEs converge at the theoretical rate of consistency, $O(1/n)$. Furthermore, the estimation error increases when the latent dimension $K_1$ (and $K_2$) is higher or the observation rate is lower. Moreover, a primary observation is that our proposed joint estimation approach consistently outperforms the separate approach in estimating both $\Tt_{1*}$ and $\Tt_{2*}$. This observation provides numerical evidence that properly incorporating the shared linear structure leads to more accurate and efficient parameter estimation.

\begin{figure}[htb!]
  \centering
  \includegraphics[width=0.85\textwidth]{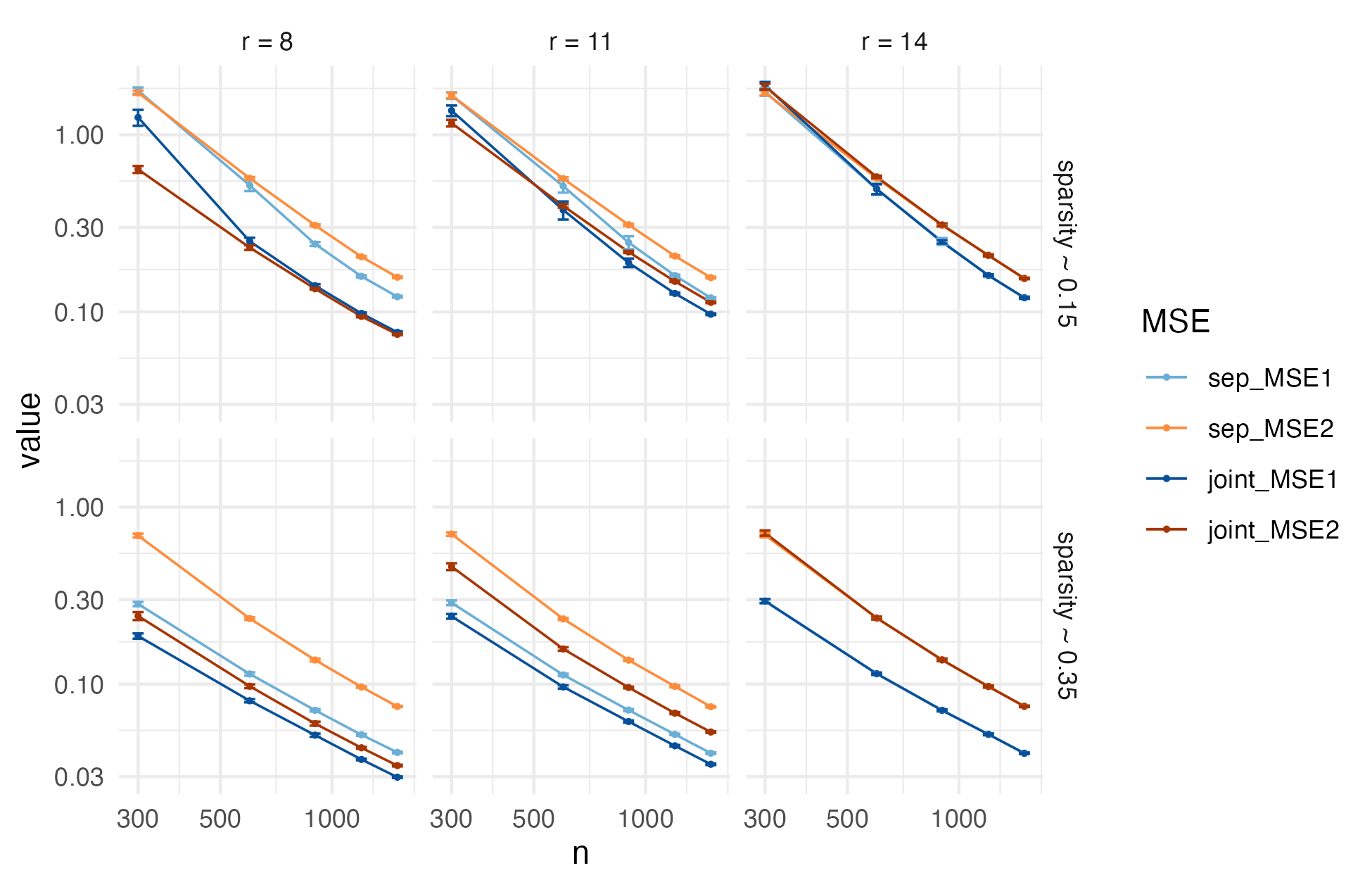}
  \caption{Mean Squared Errors (MSEs) for $\Tt_{1*}$ and $\Tt_{2*}$ as a function of $n$. Panels show results for varying latent dimensions ($r$) and observation rates (sparsity), with a fixed $K_1=6$. Separate and joint estimations are represented by light- and dark-colored lines, respectively. Both axes are on a $\log_{10}$-scale. Error bars represent $\pm 1$ standard error over $20$ data realizations. }\label{fig:asym_fix_K1}  
\end{figure}

In the second experiment, we fix $K_1 = 6$ and vary $r\in\{8, 11, 14\}$, representing a large, moderate, and small amount of shared information between the data matrix and its missingness pattern, respectively. The results are presented in Figure~\ref{fig:asym_fix_K1}. As in the first experiment, the results are consistent with our theoretical analysis. The key finding concerns the relationship between the benefit of joint modeling and the amount of shared information. When there is no shared structure ($r=14$), the estimation errors of separate and joint estimations are virtually identical. As the amount of shared information increases (i.e., as $r$ decreases), the performance of separate estimation remains unchanged as expected, and the improvement offered by joint estimation becomes more substantial. The benefit of joint modeling is maximized when $r=8$, in which case $\Tt_{1*}$ and $\Tt_{2*}$ share the same column and row spaces.

\subsection{Finite-Sample Comparison with Other Methods}

We compare the separate and joint estimation approaches with the \textsc{Soft-Impute} algorithm \citep{mazumder10a} and an IPW-based approach \citep{mao2021matrix}.
\textsc{Soft-Impute} minimizes the least-squares loss \eqref{eq:sep_loss1} over $\Tt_1$ with a nuclear-norm penalty, without modeling the missingness pattern. The IPW-based approach first fits a nonlinear low-rank model on $Z$, then estimates $\Tt_{1*}$ by minimizing a nuclear-norm penalized least-squares loss weighted by the estimated inverse probabilities. 
For both competing methods, the order of the tuning parameter is set according to the theoretical results in \citet{klopp2015} and \citet{wang2021matrix}, respectively. We enumerate the leading penalty constants over a small grid of candidate values, rather than select them using cross-validation. For the separate and joint estimations, the latent dimensions $K_1$, $K_2$, $r$, and $s$ are treated as known.

\begin{figure}[htb!]
\centering
\includegraphics[width=0.85\textwidth]{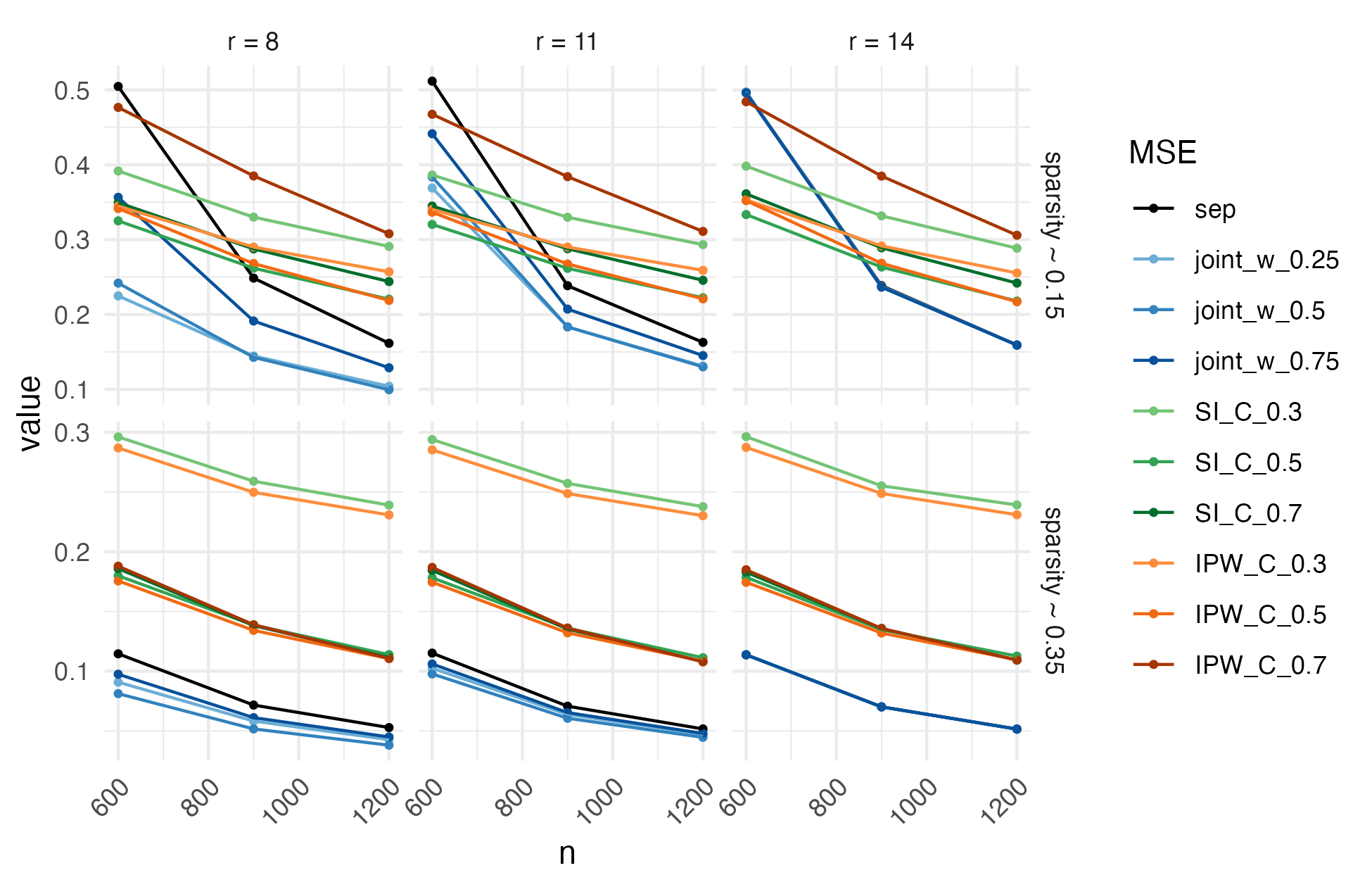}
\caption{Mean Squared Errors (MSEs) for estimating $\Tt_{1*}$ as a function of $n$. Panels show results for varying latent dimensions ($r$) and observation rates (sparsity), with a fixed $K_1=6$. \textsc{Soft-Impute} and the IPW-based approach are labeled by the prefixes \texttt{SI\_} and \texttt{IPW\_}, respectively. Suffixes and color shading denote different tuning parameter choices, including the penalty constant (\texttt{C}) and the weight for joint estimation (\texttt{w}). Reported errors are averaged over 20 data realizations.
}\label{fig:cons} 
\end{figure}

In the experiment, we vary $n\in\{600, 900, 1200\}$, and for the latent dimensions, we fix $K_1 = 6$ and vary $r\in\{8, 11, 14\}$ to simulate different amounts of shared information between the data matrix and its missingness pattern. Figure~\ref{fig:cons} reports the MSEs for estimating $\Tt_{1*}$. 
With a sufficient amount of data---from either a high observation rate ($35\%$) or a large $n$---both separate and joint estimations outperform the two competing methods. This is likely due to the bias introduced by the nuclear-norm penalty and the fact that our methods utilize the true latent dimensions. 
With a low observation rate ($15\%$) and small $n$, the performance of both separate and joint estimations degrades. This aligns with the finding that matrix completion methods relying solely on low-rank constraints tend to overfit limited data \citep{mnih2007probabilistic}. 
Nonetheless, when information is shared between $Y$ and $Z$ through the shared linear structure, our joint estimation approach effectively mitigates this overfitting. As $r$ becomes smaller, the benefit of joint modeling becomes more substantial. 
Finally, with properly chosen tuning parameters, the IPW-based approach only slightly outperforms \textsc{Soft-Impute} given sufficient data, and offers no benefit otherwise. This limited advantage reveals the drawback of IPW when the estimation of observation probabilities suffers from large variation. 

\subsection{Selection Consistency of the Criterion-Based Procedure}

In this subsection, we set the true latent dimensions to $K_{1*}=6$, $K_{2*}=8$, $r_{*}=11$ and $s_{*}=11$, and select $K_1$, $K_2$, $r$ and $s$ from their corresponding candidate sets using the criterion-based selection procedure proposed in Section~\ref{sec:model_selection}. The candidate set for each latent dimension $d \in \{K_1, K_2, r, s\}$ is the set of integers $\{d_*-3, d_*-2, \dots, d_*+3\}$, where $d_*$ is its true value. When determining $r$ and $s$, $K_{1*}$ and $K_{2*}$ are treated as known. We vary $n$ from $600$ to $2100$ in increments of $300$.

We first describe how we investigate the asymptotic behavior of the information criteria, using the selection of $K_m$ ($m=1,2$) as an example. 
Recall that the constant $\rho$ in criterion~\eqref{eq:IC_sep} controls the strength of the penalty for model complexity. Instead of presenting the rank selection consistency under a fixed $\rho$, we plot its valid interval $(\rho_-, \rho_+)$ under various experimental settings. Specifically, $\rho_-$ represents the minimum penalty strength required to prevent over-estimation of $K_m$, while $\rho_+$ is the maximum threshold before under-estimation occurs. With any $\rho\in(\rho_-, \rho_+)$, the true latent dimension $K_{m*}$ will be correctly selected. By Theorem~\ref{thm:IC_joint}, the width of $(\rho_-, \rho_+)$ should increase as $n$ increases, making the choice of $\rho$ asymptotically inconsequential. 
This provides a more comprehensive view of the performance of the criterion, such as its robustness against the choice of $\rho$ under different sample sizes. 
A similar argument applies to the selection of $r$ and $s$, with the constant $\tilde{\rho}$ in criterion~\eqref{eq:IC_joint} and the corresponding interval $(\tilde{\rho}_-, \tilde{\rho}_+)$.

\begin{figure}[htb!]
\centering
\includegraphics[width=0.95\textwidth]{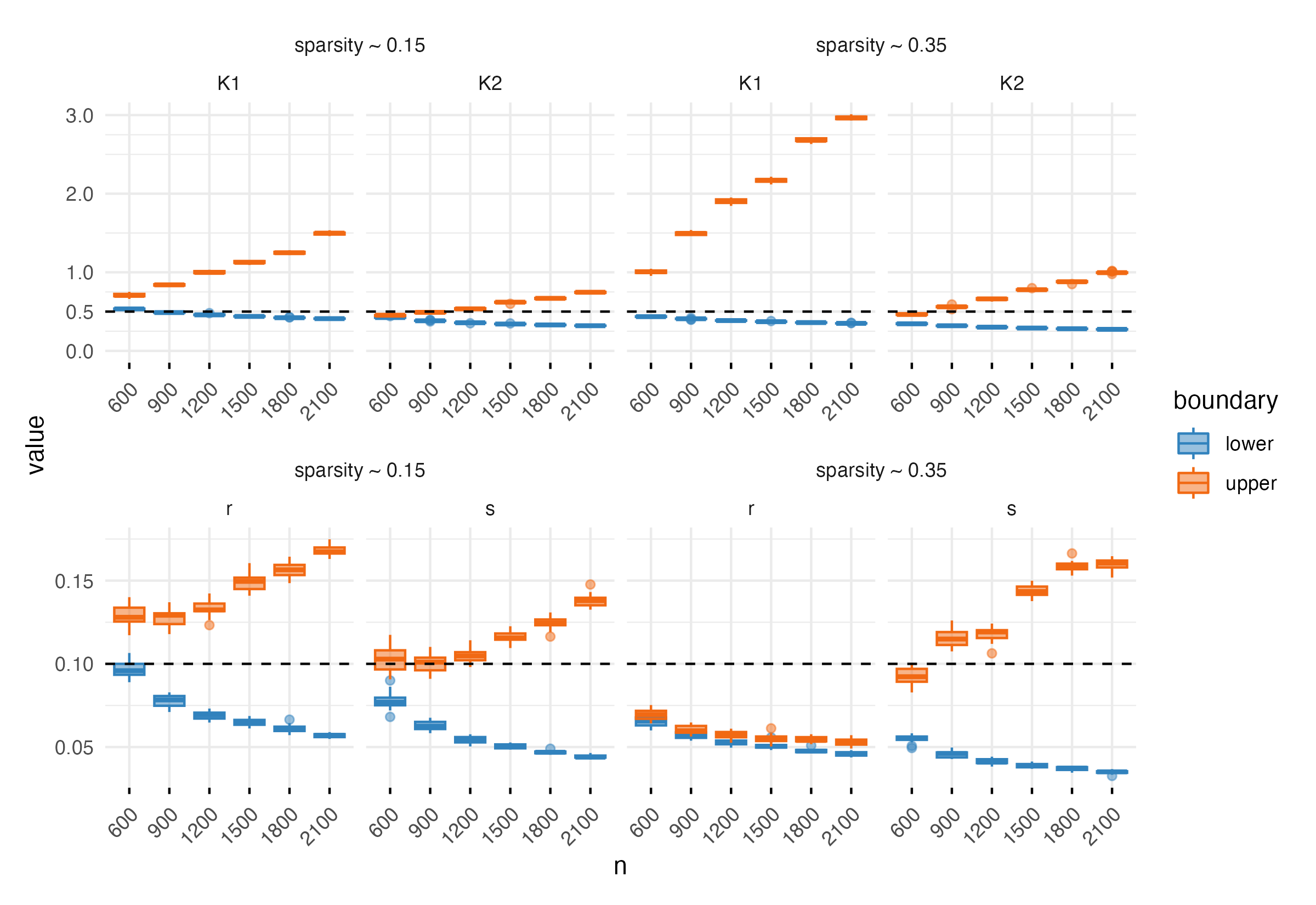}
\caption{
Valid intervals for the penalty constants $\rho$ (top row, for $K_1, K_2$) and $\tilde{\rho}$ (bottom row, for $r, s$) in criteria~\eqref{eq:IC_sep} and \eqref{eq:IC_joint}. Boxplots show the distributions of the lower and upper boundaries of these intervals over $20$ data realizations. The separation between the boxes for a given $n$ represents the range of values that correctly identify the true latent dimension in all realizations. Dashed lines indicate $\rho = 0.5$ and $\tilde{\rho}=0.1$. Panels correspond to different latent dimensions and observation rates (sparsity).
}\label{fig:rank} 
\end{figure}

Figure~\ref{fig:rank} presents boxplots of the lower and upper boundaries of the intervals $(\rho_-, \rho_+)$ and $(\tilde{\rho}_-, \tilde{\rho}_+)$, computed over $20$ data realizations.
The results yield the following observations.
Across all four latent dimensions and both observation rates, the intervals for valid $\rho$ and $\tilde{\rho}$ widen as $n$ increases, demonstrating the rank selection consistency of the proposed criterion-based procedure, where the specific choice of constants $\rho$ and $\tilde{\rho}$ becomes asymptotically negligible. 
Specifically, with $\rho = 0.5$ and $\tilde{\rho}=0.1$, our procedure correctly identifies the true latent dimensions with a moderate sample size in most cases. A notable exception is the selection of $r$ at an observation rate of $35\%$. In this case, the upper boundaries of the intervals are systematically lower than those for other latent dimensions, although the widening trend persists. 
This is likely due to the randomness in generating the true parameters, especially $\{ \L_{1*}, \R_{1*}, \L_{2*}, \R_{2*} \}$, as the results for each observation rate are based on a single instance of $\{ \L_{1*}, \R_{1*}, \L_{2*}, \R_{2*} \}$ shared across all sample sizes. 
Finally, comparing the widths of intervals in the first and second rows, we note that the proposed procedure is generally more robust to the choice of $\rho$ than to that of $\tilde{\rho}$.

\section{Real-World Data Analysis}\label{sec:real_data}

The MovieLens 100K Dataset \citep{harper2015movielens} contains $100{,}000$ anonymous ratings on a five-star scale from $n = 943$ users for $1{,}682$ movies. 
The overall observation rate is about $0.06$, while it varies across users and movies, indicating the potential of utilizing information from the missingness pattern.
In the dataset, each user has rated at least $20$ movies, and we remove movies receiving fewer than $10$ ratings.
This results in a total of $97{,}953$ ratings for $p=1{,}152$ movies, and the overall observation rate increases to about $0.09$. 
The objective on this dataset is to predict users' ratings for movies they have never seen, i.e., the unobserved entries in the user-movie rating matrix. In this section, we compare our proposed joint low-rank approach against several state-of-the-art methods on this task.

Recall that $Y\in\mathbb{R}^{n\times p}$ is the partially observed data (rating) matrix, and $Z\in\{0,1\}^{n\times p}$ is its missingness pattern. 
The original dataset already includes a $5$-fold split of the observed ratings for benchmarking and evaluation. We reserve one fold as the test set at a time, and evaluate the competing methods on the $5$ test sets. 
When fitting our joint model, we mask two sets of entries in $Z$: (1) all user-movie pairs in the test set (where $Z_{ij}=1$) to avoid using the observation status of the test entries, and (2) a random sample of $20\%$ of the unobserved pairs (where $Z_{ij}=0$) to prevent selection bias induced by the removal of (1). 
Although $Z$ also contains missingness after masking these entries, 
both the separate and joint estimation approaches can be easily modified to allow missingness in $Z$. Let $\Omega\in\{0,1\}^{n\times p}$ be the indicator of the test set.
Denoting by $\hat{\Tt}_1$ the estimated target matrix, we use the prediction root mean square error (RMSE) as the evaluation metric for prediction accuracy. It is formally defined as
$$
\mathrm{RMSE}(\hat{\Tt}_1) := \sqrt{\frac{\|\Omega \circ (Y - \hat{\Tt}_1)\|_{\fn}^2}{\|\Omega\|_{\fn}^2}}.
$$

\begin{figure}[htb!]
\centering
\includegraphics[width=\textwidth]{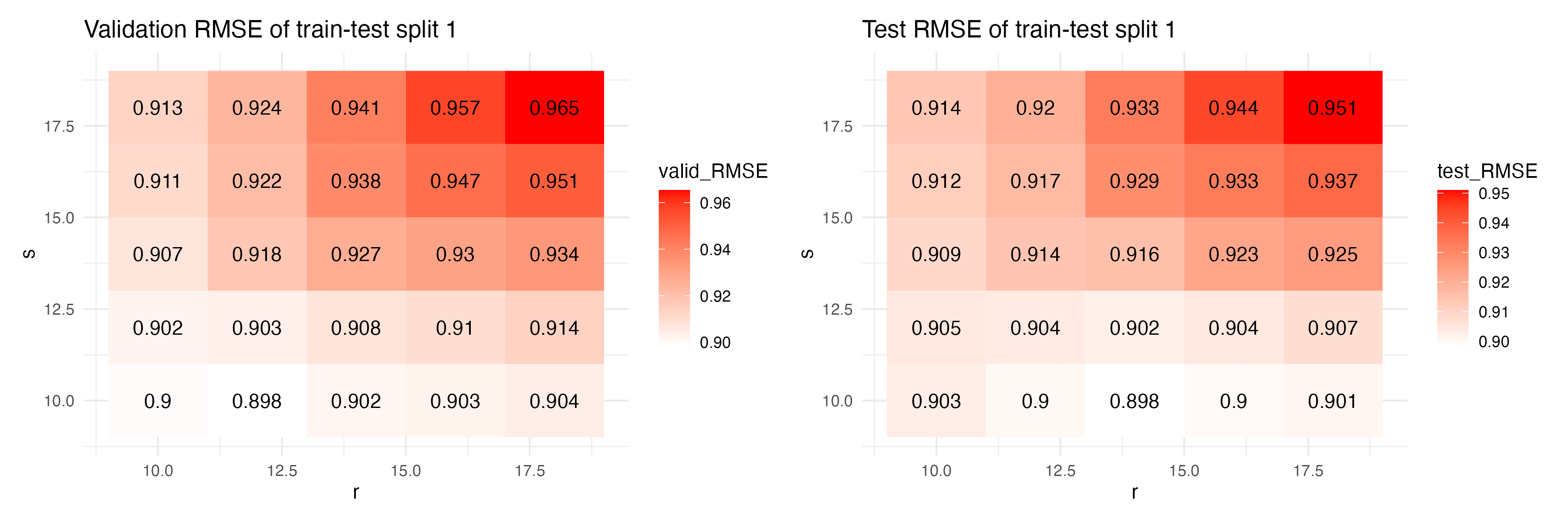}
\caption{
Heatmaps of the average validation (left) and test (right) RMSEs of the joint model under various latent dimensions $r$ and $s$. Results are shown for the first train-test split with the weight parameter fixed at $w=0.7$. 
}\label{fig:grid} 
\end{figure}

The proposed joint low-rank model is fitted using the estimation procedure detailed in Section~\ref{sec:estimation}. When separately estimating the two sub-models, the latent dimensions $K_1$ and $K_2$ are determined via $5$-fold cross-validation, minimizing the average prediction RMSE on the validation sets. We select $K_1 = 8$ and $K_2=10$ for all $5$ train-test splits. 
In the joint estimation, we need to further specify the weight parameter $w$ and the latent dimensions $r$ and $s$, which characterize the amount of shared information. Recall that $r$ and $s$ are required to satisfy the following constraints,
$$
10=K_1\vee K_2 \leq r, s \leq \ K_1 + K_2=18.
$$
We perform a grid search over $w\in\{0.3, 0.5, 0.7, 0.9\}$ and $r,s\in\{10, 12, 14, 16, 18\}$, and determine $w$, $r$, and $s$ using $5$-fold cross-validation. 
This consistently yields $w=0.7$, $r=12$, and $s=10$ across all $5$ train-test splits. In Figure~\ref{fig:grid}, we show the average validation RMSEs and test RMSEs on the first train-test split with $w=0.7$ and all combinations of $r$ and $s$. We can see that $r=12$ and $s=10$ yield the lowest average validation RMSE, and the RMSE increases as $r$ and $s$ deviate from their optimal values. This demonstrates that modeling the association between the data matrix and its missingness pattern can improve prediction performance, and that the flexibility of our modeling framework plays an important role in maximizing this benefit. 

\begin{figure}[htb!]
\centering
\includegraphics[width=0.7\textwidth]{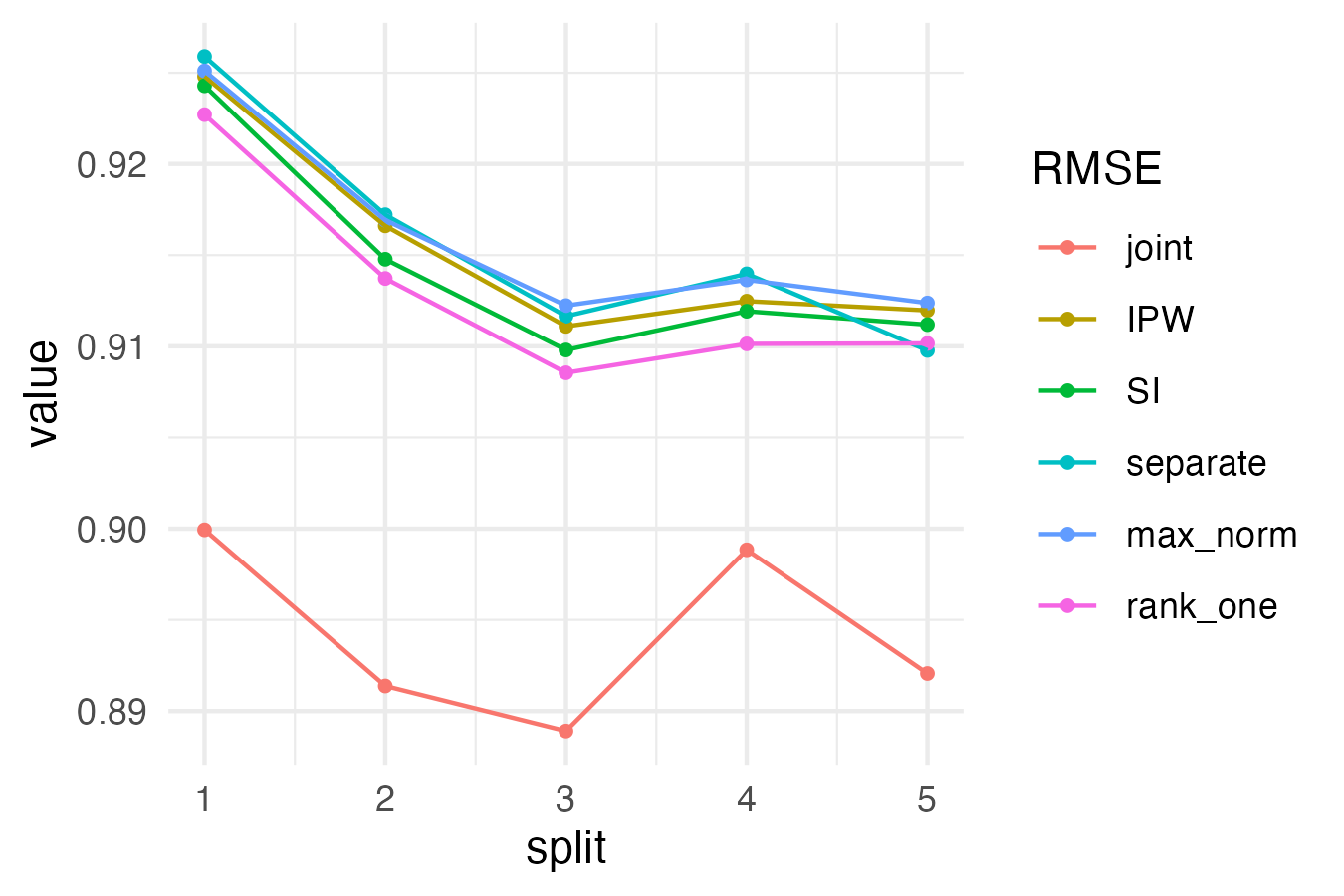}
\caption{
Comparison of test prediction RMSEs for all competing methods across $5$ train-test splits of the MovieLens 100K dataset. 
}\label{fig:real} 
\end{figure}

We compare the proposed joint low-rank approach with (1) \texttt{separate}: the separate estimation approach; (2) \texttt{IPW}: the IPW-based approach \citep{mao2021matrix}; (3) \texttt{SI}: the \textsc{Soft-Impute} algorithm \citep{mazumder10a}; (4) \texttt{max\_norm}: the max-norm regularization approach \citep{cai2016matrix, fang2018max};
and (5) \texttt{rank\_one}: the weighted nuclear-norm regularization approach assuming a rank-one observation probability matrix \citep{srebro2010collaborative, foygel2011learning, negahban2012restricted}. 
All tuning parameters are selected using cross-validation. The test prediction RMSEs on all $5$ train-test splits are reported in Figure~\ref{fig:real}. Our joint low-rank approach consistently outperforms other competing methods across all $5$ test sets, achieving a reduction in test RMSE of more than $0.02$. This improvement is substantial given the large size of this dataset.
All remaining methods have comparable performance, with the weighted nuclear-norm regularization approach (\texttt{rank\_one}) slightly outperforming the others. The IPW-based approach fails to leverage the informative missingness and even performs worse than the vanilla \textsc{Soft-Impute}, revealing its limitation in low-observation-rate settings. In summary, the results show that our joint low-rank model with the shared linear structure can successfully extract information from the missingness pattern and improve prediction performance on the MovieLens 100K dataset.

\subsection{Using Fitted Embeddings in Downstream Tasks}\label{sec:downstream}

Beyond the improved prediction accuracy, another advantage of our joint low-rank approach is that it provides the unified embeddings, $\F$ and $\Ld$, as comprehensive and compact representations of the users and movies, which can be used in downstream tasks. 
In user preference datasets, the user profiles are an important source of information, yet user demographic data may not always be available due to privacy concerns or data collection overhead. 
We hypothesize that the unified embeddings implicitly capture meaningful user characteristics beyond the preferences themselves, and thus can be useful in reconstructing user profiles.

\begin{figure}[htbp]
\centering
\includegraphics[width=0.75\textwidth]{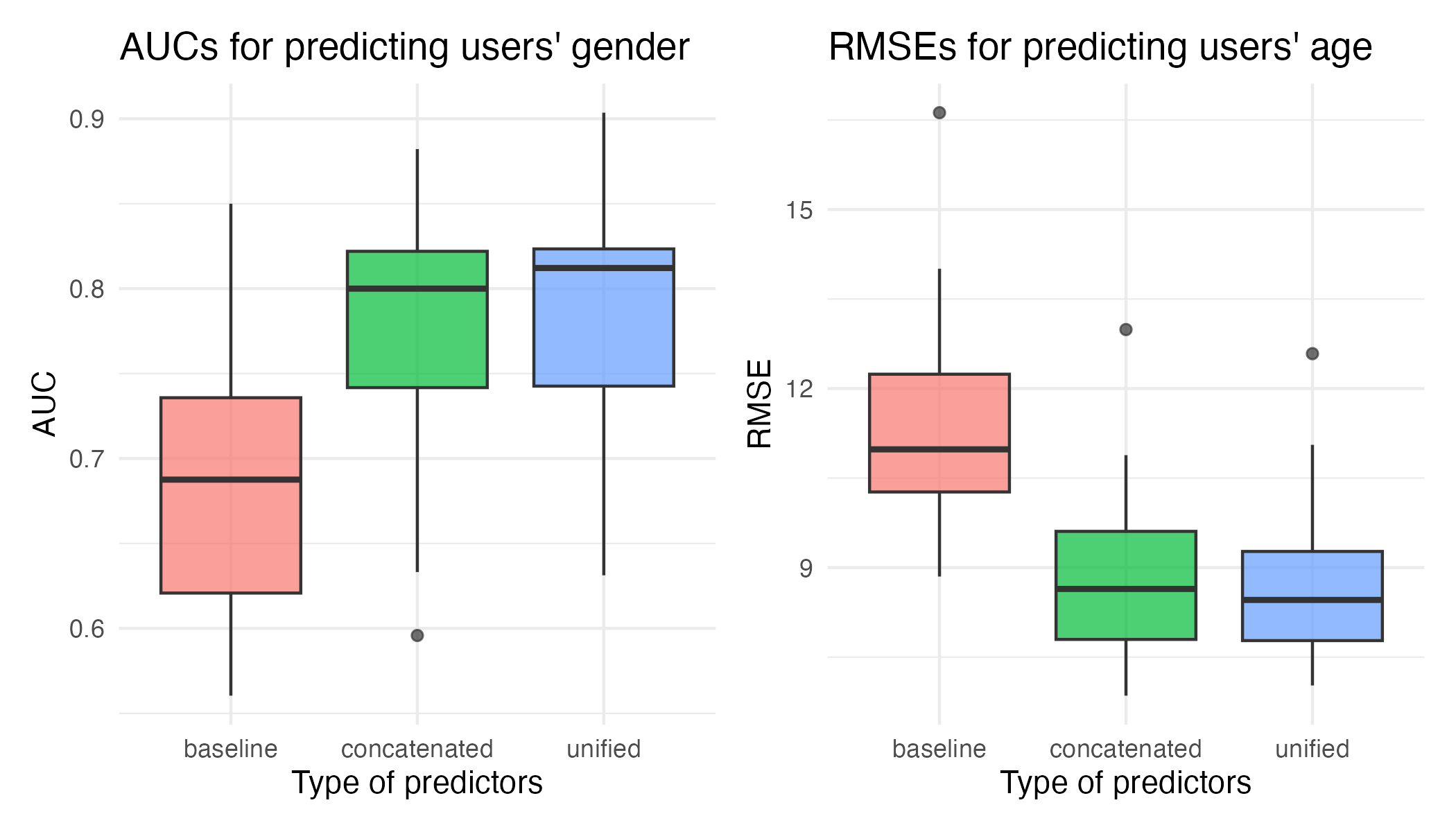}
\caption{
Performance of different user embeddings on downstream prediction tasks. The boxplots show the distribution of AUCs for gender prediction (left) and RMSEs for age prediction (right) over $20$ test folds. The comparison includes the baseline ($\hF_1$), concatenated ($(\hF_1\ \hF_2)$), and unified ($\hF$) embeddings.
}\label{fig:ds} 
\end{figure}

To validate this hypothesis, we use the estimated unified embeddings of the users, $\hF$, as the predictors to predict their gender and age. 
Specifically, with the selected dimensions $K_1=8$, $K_2 = 10$, $r = 12$, $s=10$, and $w=0.7$, we refit the joint low-rank model using the entire dataset to obtain the refitted embeddings $\hF$. For comparison, we also consider using $\hF_1$ from the separate estimation as the baseline embeddings, as well as the concatenated embeddings $(\hF_1\ \hF_2)\in\mathbb{R}^{n\times (K_1 + K_2)}$. The concatenated embeddings ideally contain at least as much information as the unified embeddings $\hF$, but they have a higher dimension and are likely noisier. 
With these three types of predictors, we fit three logistic regression models to predict the gender of the users, and fit three linear models to predict their age. Note that although the embeddings are not identifiable in the low-rank models, the transformations do not change the columns spaces of them, thus the identifiability issues are inconsequential to the predictions. 
To evaluate the performance on these downstream tasks, we partition the users into $20$ folds and record the test AUCs and RMSEs for predicting gender and age, respectively. The evaluation metrics are reported in Figure~\ref{fig:ds}. We can see that for both tasks, the unified embeddings yield significantly better prediction performance than the baseline embeddings, and slightly outperform the concatenated embeddings. 
This confirms that the estimated unified embeddings are interpretable and successfully encode intrinsic user heterogeneity.

\section{Discussion}\label{sec:discussion}

In this paper, we propose a joint low-rank modeling framework that jointly models the data matrix and its missingness pattern in order to incorporate the informative missingness for noisy matrix completion. In particular, we assume that the two modes of data follow a pair of low-rank models whose parameters are linked by a novel shared linear structure. Our framework can flexibly model various associations between the data matrix and its missingness pattern, and it provides a set of comprehensive and compact representations for the row and column items of the data matrix. 
Scalable estimation algorithms with theoretical guarantees are provided for the proposed model. We note that our modeling framework can be generalized to model the dependency between $Y_{ij}$ and $Z_{ij}$ by treating the unified embeddings $\f_i$'s and $\ld_j$'s as latent variables instead of fixed parameters. 
Further treating $\L_1$ and $\L_2$ as latent variables would allow us to model the peer effect, i.e., statistical dependency across users.

There are a few directions we may continue to explore in the future. 
First, due to the technical challenge encountered in estimating the binary low-rank model on $Z$, our theoretical results rely on the condition $\pi_- = \Omega((n\wedge p)^{-1/2})$. We may investigate whether this can be relaxed to align with assumptions in the standard matrix completion literature.
Second, our local convergence analysis of projected gradient descent assumes that the latent dimensions are correctly specified. A natural and important extension is to analyze the over-specified case, which is highly relevant for practical model selection.
Finally, we may study how to conduct statistical inference for the target low-rank matrix so that the uncertainty in the prediction can be quantified.


\acks{}



\vskip 0.2in
\bibliography{reference}

\end{document}